\documentstyle[12pt]{article}
\begin{document}
\begin{flushright}
hep-th/0212240\\
SNB/December/2002
\end{flushright}

\vskip 1.7cm

\begin{center}
{\bf { \large GAUGE TRANSFORMATIONS, BRST COHOMOLOGY \\
AND WIGNER'S LITTLE GROUP}}

\vskip 2.5cm

{\bf R.P.Malik}
\footnote{ E-mail address: malik@boson.bose.res.in  }\\
{\it S. N. Bose National Centre for Basic Sciences,} \\
{\it Block-JD, Sector-III, Salt Lake, Calcutta-700 098, India} \\

\vskip 2cm

\end{center}

\noindent
{\bf Abstract:}
We discuss the (dual-)gauge transformations and BRST cohomology for the
two $(1 + 1)$-dimensional (2D) free Abelian one-form and four
($3 + 1)$-dimensional (4D) free Abelian 2-form gauge theories by exploiting the
(co-)BRST symmetries (and their corresponding generators) for the Lagrangian
densities of these theories. For the 4D free 2-form gauge theory, we show that
the changes on the antisymmetric polarization tensor $e^{\mu\nu} (k)$
due to (i) the (dual-)gauge transformations corresponding to the internal
symmetry group, and (ii) the translation subgroup $T(2)$ of the Wigner's
little group, are connected with each-other for the specific
relationships among the parameters of these transformation groups. In
the language of BRST cohomology defined w.r.t. the conserved and nilpotent
(co-)BRST charges, the (dual-)gauge transformed states turn out to be the sum
of the original state and the (co-)BRST exact states. We comment on (i)
the quasi-topological nature of the 4D free 2-form gauge theory from the
degrees of freedom count on $e^{\mu\nu} (k)$, and (ii) the Wigner's little
group and the BRST cohomology for the 2D one-form gauge theory
{\it vis-{\`a}-vis} our analysis for the 4D 2-form gauge theory.\\



$~~~~~~~~~~$ Hodge decomposition theorem; Wigner's little group

%\vskip 1cm

\baselineskip=16pt

\newpage

\noindent
{\bf 1 Introduction}\\

\noindent
It is a fairly well-established fact that the local gauge invariant theories
govern the three (out of four) fundamental interactions of nature.
In particular, the one-form (non-)Abelian gauge theories (which are at
the heart of the standard model) have been verified and tested
experimentally with stunning degree of
precision. A couple of key features of these
gauge theories are (i) the existence of the first-class constraints
(in the language of Dirac's classification scheme [1,2]) on them
that turn out to generate the local gauge transformations, and
(ii) the interaction between gauge fields and matter fields is always
dictated by the requirement of local gauge invariance,
because of which, the gauge fields of the theory couple to the Noether
conserved current (constructed by the matter fields). The latter are derived
due to the presence of the continuous global
gauge symmetry in the theory. For some of the free as well as interacting
gauge theories, however, there exists a discrete symmetry
transformation
(for the Lagrangian density of these gauge theories) that corresponds
to a certain specific kind of ``duality'' in the theory.  This duality is
found to be responsible for the existence of (i) a local dual-gauge symmetry
transformation in the theory, and (ii) the analogue of the
Hodge duality $*$ operation of differential geometry defined on a compact
spacetime manifold without a boundary (see, e.g., [3-7]).
Such (non-)interacting duality invariant
gauge theories are found to provide a set of
tractable field theoretical models for the Hodge theory. In this context,
we can mention many interesting (gauge invariant)
field theoretical models such as: (i) the free $(1 + 1)$-dimensional (2D)
Abelian gauge theory [8-10], (ii) the
interacting 2D Abelian gauge theory where there
is an interaction between the $U(1)$ gauge field and the Dirac fields [11,12],
(iii) the self-interacting 2D non-Abelian gauge theory where there is no
interaction between matter fields and the non-Abelian gauge field [10,13], and
(iv) the free Abelian 2-from gauge theory in $(3 + 1)$-dimensions (4D) [14].

For the covariant canonical quantization of any gauge
theory (endowed with the first-class constraints),
the Becchi-Rouet-Stora-Tyutin (BRST) formalism is one of the
most elegant methods where the unitarity and ``quantum'' gauge
(i.e. the BRST) invariance are respected together at any arbitrary order
of perturbation theory. In this formalism, the usual local ``classical''
gauge transformations of the original
gauge theory are traded with the ``quantum'' gauge transformations
which are popularly known as the BRST transformations. One of the key
features of these transformations is the nilpotency property.
In fact, these symmetry transformations are generated by a conserved
($\dot Q_{b} = 0$) and nilpotent ($Q_{b}^2 = 0$) BRST charge $Q_{b}$.
As it turns out, the first-class constraints of the original gauge theory
are found to be encoded in the physicality condition $Q_{b} \; |phy> = 0$
in a subtle way. In fact, the condition $Q_{b} |phy> = 0$ implies that
the physical states ($|phy>$) belong to the subspace of the total
quantum Hilbert space of states (QHSS) that are annihilated by the BRST charge
(or equivalently by the operator form of the  first-class constraints of
the original gauge theory). The nilpotency of the BRST charge
and the physicality condition allow one to define the BRST cohomology
where two quantum states ($ |phy>^\prime$ and $|phy>$)
are said to belong to the same cohomology class w.r.t. the BRST charge $Q_{b}$
if they differ by a BRST exact state (i.e. $ |phy>^\prime = |phy>
+ Q_{b}\;|\Omega>$ for an arbitrary state $|\Omega>$ in the
quantum Hilbert space). The BRST charge $Q_b$ turns out to be the analogue
of the exterior derivative $d$ (with $d^2 = 0$) of differential geometry.
Because of the presence of the duality, however, it
is found that a local dual-gauge transformation also exists
for these gauge theories. The latter transformation too
can be traded with a
nilpotent dual(co-)BRST symmetry transformations which are generated by
a conserved $(\dot Q_{d} = 0)$ and nilpotent ($Q_{d}^2 = 0$) co-BRST
charge $Q_{d}$. Analogous to the cohomology w.r.t. $Q_{b}$, a co-cohomology
can be defined w.r.t. the co-BRST charge $Q_{d}$.  This charge turns out to
correspond to the co-exterior derivative $\delta = \pm * d *$
(with $\delta^2 = 0$) of the differential geometry. As argued earlier, the
discrete symmetry transformation of the theory (corresponding to the
so-called ``duality'') turns out to be the analogue of the $*$ operation of
the differential geometry (cf. (3.5) below). In particular,
the discrete symmetry transformations on the ghost fields
are exploited to define the anti-BRST and anti-co-BRST symmetries that
are generated by the conserved and
nilpotent ($Q_{ab}^2 = Q_{ad}^2 = 0$) anti-BRST
($Q_{ab}$) and anti-co-BRST ($Q_{ad})$ charges.
Furthermore, the anti-commutator of the above nilpotent symmetries
generates a bosonic symmetry transformation. The generator
of this bosonic symmetry transformation $Q_{w}$ turns out to be the
analogue of the Laplacian operator $\Delta$ (i.e $\Delta = (d + \delta)^2
= \{ d, \delta \}$) of the differential geometry. Thus, all the de Rham
cohomological operators $(d, \delta, \Delta)$ find their analogues in the
language of some local, covariant and continuous symmetry transformations
(and their generators) for the Lagrangian density of
such a duality invariant gauge theory.
Yet another symmetry in the theory is the ghost scale symmetry in which the
(anti-)ghost fields transform by a scale transformation and the rest of the
physical fields do not transform at all. This continuous symmetry
transformation is generated by a conserved ghost charge $Q_{g}$. Thus, we
have six local and conserved charges in the theory.

Having defined, described and discussed about the above conserved charges,
now the stage is set for the statement of the celebrated Hodge decomposition
theorem (HDT) in the QHSS. In fact, any
arbitrary
state $|\psi>_{n}$ with the ghost number $n$ (i.e. $ i Q_{g} |\psi>_n = n \;
|\psi>_n $) in the QHSS
can be expressed as the sum of a harmonic state $|\omega>_n$
(with $Q_{w} |\omega>_n = 0, Q_{b} |\omega>_n = 0, Q_{d} |\omega>_n = 0$),
a BRST exact state $Q_{b} |\phi>_{n-1}$ and a BRST co-exact state
$Q_{d} |\theta>_{n + 1}$ as
\footnote{In the language of differential geometry defined on a
compact manifold without a boundary, the HDT states that any arbitrary form
$f_{n}$ of degree $n$ can be decomposed into a harmonic form $h_n$
(with $ \Delta h_n = 0, d h_n = 0, \delta h_n = 0$), an exact form
$d e_{n-1}$ and a co-exact form $\delta c_{n+1}$ as:
$f_{n} = h_{n} + d\;e_{n-1} + \delta\; c_{n+1}$ where $(d, \delta, \Delta)$
form the de Rham cohomological set of operators on the manifold.}
$$
\begin{array}{lcl}
|\psi>_{n} = |\omega>_n + Q_{b}\; |\phi>_{n -1} + Q_{d}\; |\theta>_{n +1}
\equiv
|\omega>_n + Q_{ad}\; |\phi>_{n -1} + Q_{ab}\; |\theta>_{n +1}.
\end{array}\eqno(1.1)
$$
Thus, it is crystal clear that there exists
 a {\it two-to-one mapping} between the
conserved charges and the cohomological operators as: $ (Q_{b}, Q_{ad})
\rightarrow d,\;(Q_{d}, Q_{ab}) \rightarrow \delta, \{Q_{b}, Q_{d} \}
= \{Q_{ab}, Q_{ad} \} = Q_{w} \rightarrow \Delta$. All the above stated
mathematical issues have been addressed and shown to be connected
with the symmetry properties of the Lagrangian densities (and their
 corresponding generators) for the field theoretical models of the gauge
theories, we have stated earlier [8-14]. In a recent
paper [15], the above cohomological properties and the quasi-topological
nature of the 4D free Abelian 2-form gauge theory have been discussed by
exploiting the (anti-)BRST and (anti-)co-BRST symmetries, their corresponding
generators, their ensuing algebraic structure and the HDT in the QHSS.

A quite different but very interesting aspect of the above discussion is
connected with the geometrical interpretations [16-21] for all the
conserved charges ($Q_{(a)b}, Q_{(a)d}, Q_{w}$) and their two-to-one
mappings with the cohomological operators $(d, \delta, \Delta)$ in the
framework of superfield formulation [22-26]. In fact, it has been
demonstrated in our earlier works on the superfield
formulation of the 2D (non-)Abelian gauge theories [16-21] that the
conserved and nilpotent (anti-)BRST and (anti-)co-BRST charges correspond
to the translational generators along the Grassmannian directions of the
four $(2 + 2)$-dimensional supermanifold. In our formulation, it is
interesting to state that (in spite of their
common connection with the translational
generators along the Grassmannian directions), one can make a clear distinction
between $Q_{(a)b}$ and $Q_{(a)d}$ because of their radically different
operations (and the corresponding ensuing effects)
on the fermionic superfields which correspond to the anti-commuting
(anti-)ghost fields of the theory (see, e.g.,
[21] for details). It has also been demonstrated that the 2D free Abelian
and self-interacting non-Abelian gauge theories belong to a new class of
topological field theories (TFTs)
in flat spacetime (see, e.g., [3]) which capture
together some of the key features of Witten and Schwarz type of
TFTs [27-29]. It is gratifying to state that for these 2D TFTs, besides
providing the geometrical origin for the conserved (and nilpotent) charges,
the geometrical interpretation for the Lagrangian density and symmetric
energy-momentum tensor has also been provided [19-21] in the language of
translations of some local (but composite) superfields along the
Grassmannian directions of the $(2 + 2)$-dimensional supermanifold.
Mathematically, the above Lagrangian density and energy momentum tensor turn
out to be the total derivatives w.r.t. the Grassmannian variables of the
$(2 + 2)$-dimensional supermanifold.

As pointed out earlier, in general, it is the first-class constraints of the
gauge theories that turn out to generate the ``classical'' local gauge
transformations for the singular Lagrangian density of the theory. In
the BRST formulation, these local symmetries of the gauge
theory are traded with the nilpotent ``quantum'' gauge
(i.e. BRST) symmetries. In a set of seminal papers [30-32]
Weinberg first and later Han et al. [33-35] demonstrated the role of the
translational subgroup $T(2)$ of the Wigner's little group in generating the
local $U(1)$ gauge transformation for the massless (one-form) gauge field of
the Maxwell theory. To be more precise, $T(2)$ (which is an Abelian invariant
subgroup of the Wigner's little group) keeps the momentum $k_\mu$ of
the light-like (massless) gauge particle invariant but it transforms
the polarization vector $e_\mu (k)$ of the Maxwell one-form
field in exactly the same
manner as the $U(1)$ gauge transformation generated by the first-class
constraints of the Maxwell theory. Thus, as far as the $U(1)$
gauge transformation for the one-form Maxwell field is concerned, there
is an interesting complication between the translational symmetries associated
with the $T(2)$ subgroup of the Wigner's (spacetime) little group and the
internal symmetries related to the
first-class constraints of the gauge theory.

The purpose of the present
paper is to demonstrate that, for the $(3 +1)$-dimensional (4D) free
Abelian 2-form gauge theory (endowed with the (dual-)gauge transformations),
the translation subgroup $T(2)$ of the Wigner's little group generates
both the gauge [36] as well as the dual-gauge transformations on the
anti-symmetric polarization tensor $e^{\mu\nu} (k)$  of the
2-form gauge field $B_{\mu\nu} (x)$ for the specific
choices of the (dual-)gauge transformation parameters in terms of the
parameters of the $T(2)$ subgroup (cf. (5.12), (5.16), (5.23), (5.24)
below). This result is also discussed in the framework of BRST formalism.
In our expositions: (i) the HDT in the QHSS,
(ii) the BRST cohomology, and (iii) the (co-)BRST symmetries
play very decisive roles. For instance, first of all, the physical state
($| phy >$) (as well as the physical vacuum) of
the theory is chosen to be the
harmonic state of the Hodge decomposed state of any arbitrary state in
the QHSS. This immediately implies that
 $Q_{b(ad)} |phy> = 0, Q_{d(ab)} |phy> = 0, Q_{w} |phy> = 0$. As a
consequence of the above requirements, we obtain certain specific type of
restrictions on the single particle quantum state (SPQS) of the 2-form
gauge field $B_{\mu\nu} (x)$. These constraints, in turn, provide very
informative relationships between the anti-symmetric polarization tensor
$e^{\mu\nu} (k)$ and the momentum 4-vector $k^\mu$ which are found to
be responsible for {\it the first step} of
reduction of the number of degrees  of freedom associated with
$e^{\mu\nu} (k)$ in 4D. At {\it the next step}, it is the nilpotent (co-)BRST
(or (dual-)gauge) symmetry transformations that dictate the reduction process
of the degrees of freedom of $e^{\mu\nu} (k)$.
At this stage, (i) we demonstrate the connection between the $T(2)$ subgroup
of the Wigner's little group and the (dual-)gauge (or (co-)BRST) symmetry
transformation group when they operate on the {\it doubly reduced} polarization
tensor $e^{\mu\nu} (k)$, and (ii) we comment on the quasi-topological
nature [15] of the 4D 2-form Abelian gauge theory in the framework of an
extended BRST formalism. Ultimately, in
the language of the BRST cohomology w.r.t. the (co-)BRST charges, the
(dual-)gauge (or (co-)BRST) transformed states (which are connected
with the transformation generated by $T(2)$ subgroup of the Wigner's
little group on the polarization tensor $e^{\mu\nu} (k)$)
turn out to be the sum of the original SPQS plus
a BRST (co-)exact state.  For the 2D
free Abelian one-form gauge theory,
it is not possible to apply the key concepts of the Wigner's little group
and its connection to the (dual-)gauge transformations. This is
because of the fact that one can gauge away both the components of the
2D polarization vector by the choice of the (dual-)gauge parameters of the
(dual-)gauge transformations. Thus, nothing remains in the theory and
this theory becomes topological in nature [8-10,19].
This fact is reflected in the matrix representation of the
Wigner's little group which becomes identity (trivial)
for the free Abelian one-form gauge theory in 2D. As a
consequence, neither momentum nor polarization vectors transform
under the Wigner's little group defined for the 2D Abelian one-form
gauge theory.
However, in the language of constraints, BRST cohomology and HDT,
one can capture mathematically as well as physically the key points of
the (dual-)gauge transformations in an elegant way (see, e.g., section 3 below).

The material of our work is organized as follows. In section 2, we define,
discuss and distinguish between the
gauge and dual-gauge symmetry transformations
for the gauge-fixed version of the Lagrangian densities of the 2D free Abelian
one-form and 4D free Abelian 2-form gauge theories. Section 3
is devoted to the discussion of (co-)BRST symmetries. We obtain the normal
mode expansion for the basic fields of both the theories and discuss about the
BRST cohomology and physicality condition in section 4. The central of our
present paper is section 5 where we establish the connection between $T(2)$
subgroup of the Wigner's little group and the (dual-)gauge transformations.
Furthermore, we express this connection in the language of BRST cohomology
and comment on the quasi-topological nature [15] of the 4D 2-form gauge theory.
Finally, in section 6, we
make some concluding remarks and point out a few future directions.\\

\noindent
{\bf 2 (Dual-)gauge transformations: Lagrangian formulation}\\

\noindent
Let us begin with the gauge-fixed
Lagrangian density ${\cal L}_{0}^{(1)}$ for a two
$(1 + 1)$-dimensional free Abelian gauge theory in the Feynman gauge
\footnote{ We adopt here the conventions and notations such that the two
dimensional flat Minkowski metric $\eta_{\mu\nu} = $ diag $(+1, -1)$ and the
totally anti-symmetric Levi-Civita tensor $\varepsilon_{\mu\nu} = -
\varepsilon^{\mu\nu}, \varepsilon_{01} = +1 = \varepsilon^{10}, (\partial \cdot
A) = \partial_{0} A_{0} - \partial_{1} A_{1}, E = - \varepsilon^{\mu\nu}
\partial_\mu A_{\nu} = \partial_{0} A_{1} - \partial_{1} A_{0} = F_{01},
\Box = \eta^{\mu\nu} \partial_\mu \partial_\nu =
(\partial_{0})^2 - (\partial_{1})^2 $. Here
the Greek indices $\mu, \nu, \kappa....= 0, 1$ stand for the spacetime
directions on the 2D manifold.} (see, e.g., [37-40])
$$
\begin{array}{lcl}
{\cal L}_{0}^{(1)} = - \frac{1}{4} F^{\mu\nu} F_{\mu\nu} - \frac{1}{2}
(\partial \cdot A)^2 \equiv \frac{1}{2} E^2 - \frac{1}{2}
(\partial \cdot A)^2,
\end{array}\eqno(2.1)
$$
where $F_{\mu\nu} = \partial_\mu A_\nu - \partial_\nu A_\mu$ is the
anti-symmetric second-rank curvature tensor (with $ F_{01} = E$ = electric
field) defined through the 2-form $F = d A = \frac{1}{2} (dx^\mu
\wedge dx^\nu) (F_{\mu\nu})$.  As is evident, this 2-form is derived by the
application of the exterior derivative $ d = dx^\mu \partial_\mu$
(with $ d^2 = 0$) on the connection one-form $A = dx^\mu A_\mu$
(where $A_\mu$ is the vector potential) and the gauge-fixing term
$(\partial \cdot A) = (- * d * A)$ is defined through the application of
the dual-exterior derivative $\delta = - * d *$ (with $ \delta^2 = 0$) on
the one-form $A$. Here, for the 2D theory,  the curvature tensor
$F_{\mu\nu}$ has no magnetic component and the $*$ operation is the
Hodge duality operation on 2D spacetime manifold. The application of the
Laplacian operator
$\Delta = (d + \delta)^2 = d \delta + \delta d$ on the one-form $A$ leads to
$\Delta A = dx^\mu \Box A_\mu$. In fact, the equation of motion ($\Box A_\mu
= 0$), emerging from the above gauge-fixed Lagrangian density, is captured by
the Laplacian operator in the sense that
it (i.e. $\Box A_\mu = 0$) can be derived from
the  requirement of the validity of the Laplace equation $\Delta A = 0$.
Together the set of geometrical operators
$(d,\delta,\Delta)$ define the de Rham cohomological
properties of the differential forms on a
compact manifold without a boundary and obey the algebra:
$ d^2 = \delta^2 = 0, \Delta = (d + \delta)^2 = \{ d , \delta \},
[ \Delta, d ] = [ \Delta, \delta ] = 0$.

It is straightforward to check that the above Lagrangian density, under
the following local $U(1)$ gauge and dual-gauge transformations
$$
\begin{array}{lcl}
A_\mu (x) &\rightarrow&  A_\mu^{(g)} (x) = A_\mu (x)
+ \partial_\mu \alpha (x), \nonumber\\
A_\mu (x) &\rightarrow& A_\mu^{(dg)} (x) =
A_\mu (x) - \varepsilon_{\mu\nu} \partial^\nu \Sigma (x),
\end{array}\eqno (2.2)
$$
remains invariant if the local infinitesimal
transformation parameters $\alpha (x)$ and
$\Sigma (x)$ are restricted to obey $\Box \alpha (x) = \Box \Sigma (x) = 0$.
At this stage, some of the key and relevant points are (i) under the gauge
and dual-gauge transformations, it is the kinetic energy term
(more precisely the electric field $E$ itself) and the gauge-fixing term
(more accurately $(\partial \cdot A)$ itself) remain invariant (i.e.
$ E \rightarrow E^{(g)} = E, (\partial \cdot A) \rightarrow
(\partial \cdot A)^{(dg)} = (\partial \cdot A))$, respectively. (ii) Exactly
identical restrictions (i.e. $\Box \alpha = \Box \Sigma = 0$)
on the transformation parameters emerge for the invariance of the gauge-fixing
term under the gauge transformation (i.e. $(\partial \cdot A)
\rightarrow (\partial \cdot A)^{(g)} = (\partial \cdot A)$
for $\;\Box \alpha = 0$) and the invariance of the curvature term (i.e. the
electric field) under the dual-gauge transformation
(i.e. $ E \rightarrow E^{(dg)} = E$ for $  \; \Box \Sigma = 0$). (iii) The
latter transformation in (2.2) are christened as the dual gauge transformation
because $(\partial \cdot A)$ and $E$ are  `Hodge dual' to
each-other from the point of view of their derivation by the application of
cohomological operators $\delta$ and $d$ on the connection one-form $A$. (iv)
It is interesting to note that, under a couple of
independent discrete symmetry transformations
$$
\begin{array}{lcl}
\partial_\mu \rightarrow \pm i \varepsilon_{\mu\nu} \partial^\nu, \;\qquad
A_\mu (x) \rightarrow A_\mu (x),
\end{array}\eqno (2.3a)
$$
$$
\begin{array}{lcl}
A_\mu (x) \rightarrow \mp i \varepsilon_{\mu\nu} A^\nu (x),\; \qquad
\partial_\mu \rightarrow \partial_\mu,
\end{array} \eqno(2.3b)
$$
the Lagrangian density (2.1) remains invariant because the kinetic energy
and the gauge-fixing terms exchange with each-other. Mathematically, this
statement can be succinctly expressed as
$$
\begin{array}{lcl}
(\partial \cdot A) \rightarrow \pm i E, \qquad  E \rightarrow
\pm i (\partial \cdot A), \qquad \;{\cal L}_{0}^{(1)} \rightarrow
{\cal L}_{0}^{(1)}.
\end{array} \eqno(2.4)
$$
A proper generalization of equations (2.3) will turn out to be the analogue
of the $*$ operation of differential geometry as we shall see later in the
framework of BRST formalism.

Equipped with our understanding of the 2D free Abelian gauge theory which sets
the backdrop, let us dwell a bit on the 4D free Abelian 2-form gauge theory
described by the gauge-fixed Lagrangian density in the Feynman gauge
\footnote{ We follow here the conventions and notations in such a way that the
flat 4D Minkowski spacetime manifold is endowed with the flat metric
$\eta_{\mu\nu} =$ diag $(+1, -1, -1, -1)$ and the totally antisymmetric
Levi-Civita tensor obeys $\varepsilon_{\mu\nu\kappa\sigma}
\varepsilon^{\mu\nu\kappa\sigma} = - 4!, \; \varepsilon_{\mu\nu\kappa\sigma}
\varepsilon^{\mu\nu\kappa\eta} = - 3! \delta^\eta_\sigma, etc., \;
\varepsilon_{0123} = + 1 = - \varepsilon^{0123}, \; \varepsilon_{0ijk} =
\varepsilon_{ijk}$. Here the Greek indices $\mu, \nu, \kappa.....= 0, 1, 2, 3$
correspond to the spacetime directions on the 4D manifold and the Latin
indices $ i, j, k, .....= 1, 2, 3$ stand only for the space directions of
the same manifold.} (see, e.g., [40,14,15])
$$
\begin{array}{lcl}
{\cal L}_{0}^{(2)} = \frac{1}{12} H^{\mu\nu\kappa} H_{\mu\nu\kappa}
+ \frac{1}{2} (\partial_\mu B^{\mu\nu}) (\partial^{\kappa} B_{\kappa\nu}),
\end{array} \eqno(2.5)
$$
where the totally anti-symmetric curvature tensor $H_{\mu\nu\kappa}
= \partial_\mu B_{\nu\kappa} + \partial_\nu B_{\kappa\mu} +
\partial_\kappa B_{\mu\nu}$ is
derived from the 3-form $ H = d B = \frac{1}{3!} (dx^\mu \wedge dx^\nu
\wedge dx^\kappa) (H_{\mu\nu\kappa})$ by application of the exterior
derivative $d$ on the connection 2-form $B = \frac{1}{2}
(dx^\mu \wedge dx^\nu) (B_{\mu\nu})$. The application of the dual exterior
derivative $\delta = - * d *$ on the 2-form $B$ leads to the definition of
the  one-form gauge-fixing term $ (\partial^\kappa B_{\kappa\mu}) (dx^\mu)
= \delta B$. The action of the Laplacian operator $\Delta$ on the 2-form
basic field $B$ yields $\Delta B = \frac{1}{2} (dx^\mu \wedge dx^\nu)
(\Box B_{\mu\nu})$ where $\Box = \eta^{\mu\nu} \partial_\mu \partial_\nu
= (\partial_{0})^2 - (\partial_{i})^2 $. Thus, the equation of motion
$\Box B_{\mu\nu} = 0$ for the above gauge-fixed Lagrangian density is
contained in the requirement that the Laplace equation $\Delta B = 0$ is
satisfied. The above
Lagrangian density remains invariant under the following gauge and dual-gauge
transformations
$$
\begin{array}{lcl}
B_{\mu\nu} &\rightarrow& B_{\mu\nu}^{(g)} =
B_{\mu\nu} + (\partial_\mu \alpha_\nu
- \partial_\nu \alpha_\mu), \nonumber\\
B_{\mu\nu} &\rightarrow& B_{\mu\nu}^{(dg)} =
B_{\mu\nu} + \varepsilon_{\mu\nu\kappa\xi}
\partial^\kappa \Sigma^\xi,
\end{array}\eqno(2.6)
$$
if the local infinitesimal parameters of the above transformations are
restricted to
satisfy $\Box \alpha_\mu - \partial_\mu (\partial \cdot \alpha) = 0, \;
\Box \Sigma_\mu - \partial_\mu (\partial \cdot \Sigma) = 0.$ Some of the
salient features, at this juncture, are (i) it is the curvature
term $H_{\mu\nu\kappa}$ (derived from $H = d B$) and the gauge-fixing
term $\partial^\mu B_{\mu\nu}$ (derived from
$\delta B = \partial^\kappa B_{\kappa\mu} dx^\mu$) that are found to remain
invariant under the gauge- and dual-gauge transformations, respectively.
(ii) The restrictions ($\Box \alpha_\mu - \partial_\mu (\partial \cdot \alpha)
= 0, \; \Box \Sigma_\mu - \partial_\mu (\partial \cdot \Sigma) = 0$)
on the local infinitesimal transformation parameters $\alpha_\mu (x)$
and $\Sigma_\mu (x) $ are identical for both the above transformations.
(iii) The above restrictions on the transformation parameters $\alpha_\mu$
and $\Sigma_\mu$ emerge when we demand (a) the invariance of the gauge-fixing
term under the gauge transformation (i.e. $
(\partial_\mu B^{\mu\nu}) \rightarrow (\partial_\mu B^{\mu\nu})^{(g)} =
(\partial_\mu B^{\mu\nu})$), and (b) the invariance  of the kinetic energy
term under the dual-gauge transformation (i.e.
$H_{\mu\nu\kappa} H^{\mu\nu\kappa} \rightarrow (H_{\mu\nu\kappa}
H^{\mu\nu\kappa})^{(dg)} = H_{\mu\nu\kappa} H^{\mu\nu\kappa}$). To be more
elaborate on this latter point, let us consider the explicit infinitesimal
version ($\delta_{D}$) of the dual-gauge transformation (2.6) applied to the
variation of the Lagrangian density
$$
\begin{array}{lcl}
\delta_{D} {\cal L}_{0}^{(2)} = \frac{1}{6} H^{\mu\nu\kappa}
\delta_{D} (H_{\mu\nu\kappa}) =
\frac{1}{2} H^{\mu\nu\kappa}
\varepsilon_{\nu\kappa\sigma\xi} \partial_\mu \partial^\sigma \Sigma^\xi,
\end{array} \eqno(2.7)
$$
where $\delta_{D} B_{\mu\nu} = \varepsilon_{\mu\nu\kappa\zeta} \partial^\kappa
\Sigma^\zeta, \;\delta_{D} (\partial_\mu B^{\mu\nu}) = 0$.
Ultimately, the expansion of the above term in explicit components,
implies the following
$$
\begin{array}{lcl}
&& \frac{1}{2} H^{\mu\nu\kappa}
\varepsilon_{\nu\kappa\sigma\xi} \partial_\mu \partial^\sigma \Sigma^\xi
= H^{\mu\nu\kappa} \varepsilon_{\mu\nu\kappa\xi}\;
\bigl [\; \Box \Sigma^\xi - \partial^\xi  (\partial \cdot \Sigma) \;\bigr ],
\nonumber\\
&&\delta_{D} {\cal L}_{0}^{(2)} = 0 \;\;\rightarrow\;\;
\Box \Sigma_\mu - \partial_\mu (\partial \cdot \Sigma) = 0.
\end{array} \eqno(2.8)
$$
It is easy to obtain the above condition by choosing $H_{\mu\nu\kappa}
= \varepsilon_{\mu\nu\kappa\xi} V^\xi$ and showing that
$$
\begin{array}{lcl}
&&\delta_{D} {\cal L}_{0}^{(2)} = 0 \;\;\rightarrow\;\; V^\mu \bigl [
\Box \Sigma_\mu - \partial_\mu (\partial \cdot \Sigma) \bigr ] = 0,
\nonumber\\
&&(\frac{1}{12} H^{\mu\nu\kappa} H_{\mu\nu\kappa}) \;\;\rightarrow \;\;
(\frac{1}{12} H^{\mu\nu\kappa} H_{\mu\nu\kappa})^{(dg)} =
(\frac{1}{12} H^{\mu\nu\kappa} H_{\mu\nu\kappa}).
\end{array} \eqno(2.9)
$$
Thus, the condition (2.8) emerges very naturally for the non-zero
axial-vector $V_\mu$. (iv) The above gauge-fixed Lagrangian density
remains invariant
(i.e. ${\cal L}_{0}^{(2)} \rightarrow {\cal L}_{0}^{(2)}$)
under the following discrete symmetry transformation
$$
\begin{array}{lcl}
B_{\mu\nu} \rightarrow \mp \frac{i}{2} \varepsilon_{\mu\nu\kappa\xi}
B^{\kappa\xi},
\end{array}\eqno (2.10)
$$
because the kinetic energy and gauge-fixing terms of the
above Lagrangian density exchange with each-other (i.e.
$ \frac{1}{12} H^{\mu\nu\kappa} H_{\mu\nu\kappa} \leftrightarrow
\frac{1}{2} (\partial_\mu B^{\mu\nu}) (\partial^\kappa B_{\kappa\nu})$)
under (2.10).

The gauge-fixed Lagrangian density (2.5) can be
generalized by introducing a massless ($\Box \phi_{1} = 0$) scalar field
$\phi_{1}$ in the gauge-fixing term, as
$$
\begin{array}{lcl}
{\cal L}_{1}^{(2)} = \frac{1}{12} H^{\mu\nu\kappa} H_{\mu\nu\kappa}
+ \frac{1}{2} (\partial_\mu B^{\mu\nu} - \partial^\nu \phi_{1})
(\partial^{\kappa} B_{\kappa\nu} - \partial_\nu \phi_{1}).
\end{array} \eqno(2.11)
$$
It is interesting to note that (i) the equation of motion ($\Box B_{\mu\nu}
= 0$) for the 2-form gauge field $B_{\mu\nu}$ remains intact in spite of the
presence of the massless scalar field
$\phi_{1}$. (ii) The scalar field $\phi_{1}$ does not transform under
the (dual-)gauge transformations discussed above. (iii) The Euler-Lagrange
equation of motion for $\phi_{1}$ field is $\Box \phi_{1} = 0$. (iv) The
kinetic energy term of the Lagrangian density (2.14) can be generalized
to include another massless ($\Box \phi_{2} = 0$) scalar field $\phi_{2}$,
 as given below
$$
\begin{array}{lcl}
{\cal L}_{2}^{(2)} &=&
 \frac{1}{2} (\partial_\mu B^{\mu\nu} - \partial^\nu \phi_{1})
(\partial^{\kappa} B_{\kappa\nu} - \partial_\nu \phi_{1})
\nonumber\\ &-& \frac{1}{2}
(\frac{1}{2} \varepsilon_{\mu\nu\kappa\zeta} \partial^\nu B^{\kappa\zeta}
- \partial_\mu \phi_{2}) (\frac{1}{2} \varepsilon^{\mu\sigma\eta\xi}
\partial_\sigma B_{\eta\xi} - \partial^\mu \phi_{2}).
\end{array} \eqno(2.12)
$$
The above Lagrangian density (2.12) leads to the following equations of motion
$$
\begin{array}{lcl}
\Box B_{\mu\nu} = 0, \qquad \Box \phi_{1} = 0, \qquad \Box \phi_{2} = 0.
\end{array} \eqno(2.13)
$$
Under the following generalization of the discrete symmetry
transformations (2.10)
$$
\begin{array}{lcl}
B_{\mu\nu} \rightarrow \mp \frac{i}{2} \varepsilon_{\mu\nu\kappa\xi}
B^{\kappa\xi}, \qquad \phi_{1} \rightarrow \pm i \phi_{2}, \qquad
\phi_{2} \rightarrow \mp i \phi_{1},
\end{array}\eqno (2.14)
$$
the Lagrangian density (2.12) remains invariant (i.e. ${\cal L}_{2}^{(2)}
\rightarrow {\cal L}_{2}^{(2)}$).\\

\noindent
{\bf 3 (Co-)BRST symmetries: on-shell nilpotent versions}\\

\noindent
The (dual-)gauge transformations for the gauge-fixed Lagrangian densities
of the 2D one-form and 4D 2-form gauge theories can be traded with the
(co-)BRST symmetries which turn out to be (the off-shell
as well as on-shell) nilpotent of order two. The BRST
invariant version of the Lagrangian density (2.1) is (see, e.g., [8-10])
$$
\begin{array}{lcl}
{\cal L}_{b}^{(1)} = - \frac{1}{4} F^{\mu\nu} F_{\mu\nu} - \frac{1}{2}
(\partial \cdot A)^2
- i \partial_\mu \bar C \partial^\mu C
\equiv \frac{1}{2} E^2 - \frac{1}{2} (\partial \cdot A)^2
- i \partial_\mu \bar C \partial^\mu C,
\end{array}\eqno(3.1)
$$
where the (anti-)ghost fields $(\bar C)C$ are anti-commuting
($\bar C C + C \bar C = 0, C^2 = \bar C^2 = 0$) in nature. It is
straightforward to check that (3.1) remains invariant under the
following on-shell ($\Box C = \Box \bar C = 0$) nilpotent
($s_{ab}^2 = s_{b}^2 = 0$) (anti-)BRST $s_{(a)b}$ transformations
\footnote{ We follow here the notations and conventions adopted by
Weinberg [37]. In fact, in its totality, the (anti-)BRST transformations
$\delta_{(A)B}$ (with $\delta_{(A)B}^2 = 0$) are product ($\delta_{(A)B}
= \chi s_{(a)b}$) of an anti-commuting
(i.e. $\chi C + C \chi = 0,\; \chi \bar C + \bar C \chi = 0,$ etc.)
spacetime independent parameter $\chi$
and $s_{(a)b}$ with $s_{(a)b}^2 = 0$.} on the basic fields of the theory
(with $s_{ab} s_{b} + s_{b} s_{ab} = 0$)
$$
\begin{array}{lcl}
&&s_{b} A_\mu = \partial_\mu C, \qquad s_{b} C = 0, \qquad s_{b} \bar C
= - i (\partial \cdot A), \nonumber\\
&&s_{ab} A_\mu = \partial_\mu \bar C, \qquad s_{ab} \bar C = 0, \qquad
s_{ab} C = + i (\partial \cdot A).
\end{array} \eqno(3.2)
$$
The salient points, at this stage, are (i) the Lagrangian density
transforms to a total derivative under (3.2). (ii) There are no restrictions
on the (anti-)ghost fields $(\bar C)C$. (iii) The gauge transformation
parameter $\alpha$ of (2.2) has been replaced
by an anti-commuting number $\chi$ and the ghost field $C$.
(iv) The restriction  $\Box \alpha = 0$ on the local infinitesimal parameter
$\alpha$ of the gauge transformation (2.2) is not required here
because the equation of motion $\Box C = 0$ takes care of it. (v) There are
two nilpotent ($s_{(a)b}^2 = 0$) (anti-)BRST transformations (3.2)
corresponding to a single gauge transformation in (2.2).

It is interesting to note that the dual-gauge transformation of the gauge
field $A_\mu (x)$ (cf. (2.2)) can be generalized to the dual(co-)BRST
symmetry transformation by the replacement $\Sigma (x) = \chi \bar C (x)$.
The ensuing on-shell ($\Box C = \Box \bar C = 0$) nilpotent $s_{(a)d}^2 = 0$
(anti-)co-BRST transformations ($s_{(a)d}$) (with $s_{ad} s_{d} + s_{d} s_{ad}
= 0$)
$$
\begin{array}{lcl}
&&s_{d} A_\mu = - \varepsilon_{\mu\nu}
\partial^\nu \bar C (x), \qquad s_{d} \bar C = 0, \qquad s_{d} C
= - i E, \nonumber\\
&&s_{ad} A_\mu = - \varepsilon_{\mu\nu} \partial^\nu C, \qquad s_{ad} C = 0,
\qquad s_{ad} \bar C = + i E,
\end{array} \eqno(3.3)
$$
leave the Lagrangian density (3.1) invariant up to a total derivative.
The (anti-)ghost fields $(\bar C)C$ are restricted in {\it no way} for the
consideration of the co-BRST symmetry transformations unlike
the local transformation parameter $\Sigma (x)$ of the dual-gauge
transformation. It is obvious that a couple
of on-shell ($\Box C = \Box \bar C = 0$) nilpotent ($s_{(a)d}^2 = 0$)
(anti-)co-BRST symmetry transformations emerge from a single dual-gauge
transformation of (2.2). Furthermore, the Lagrangian
density (3.1) remains invariant under the following two independent
discrete transformations
$$
\begin{array}{lcl}
\partial_\mu \rightarrow \pm i \varepsilon_{\mu\nu} \partial^\nu,\; \qquad
A_\mu (x) \rightarrow A_\mu (x),
\qquad C (x) \rightarrow \pm i \bar C (x),
\qquad \bar C (x) \rightarrow \pm i C (x),
\end{array}\eqno (3.4a)
$$
$$
\begin{array}{lcl}
A_\mu (x) \rightarrow \mp i \varepsilon_{\mu\nu} A^\nu (x), \;\qquad
\partial_\mu \rightarrow \partial_\mu,
\qquad C (x) \rightarrow \pm i \bar C (x),
\qquad \bar C (x) \rightarrow \pm i C (x),
\end{array} \eqno(3.4b)
$$
which are nothing but the generalization of transformations (2.3)
to include the transformations on the (anti-)ghost fields.
Now the stage is set to provide the meaning for the Hodge duality
$*$ operation of differential geometry in the language of BRST
type symmetries. It can be checked clearly that, for any generic field
$\Phi= A_\mu, C, \bar C$ of the theory, the following relationship
(see, e.g., [10])
$$
\begin{array}{lcl}
s_{(a)d} \Phi = \pm \;* \; s_{(a)b} \; *\; \Phi,
\end{array} \eqno(3.5)
$$
is valid. Here the Hodge duality $*$ operation
in the above equation corresponds to the discrete symmetry
transformations (3.4). The $(+)-$ signs in the above equation are
dictated  by the similar signs that appear when two successive
Hodge duality $*$ operations
are applied on the generic field $\Phi = A_\mu, C, \bar C$ as expressed below
[41,10]
$$
\begin{array}{lcl}
*\; (\; *\;) \Phi = \pm \; \Phi.
\end{array} \eqno(3.6)
$$
It will be noted that the signs on the r.h.s. of the above equation, in
general, may differ for the discrete transformations in (3.4a) and (3.4b)
(see, e.g., [10]).
It is obvious that the relationship in (3.5) is reminiscent of the connection
between the (dual-)exterior derivatives $(\delta)d$ which are related to
each-other (i.e. $\delta = \pm * d *$). In the realm of
differential geometry, the $(+)-$ signs are dictated by the
dimensionality of the manifold on which $d, \delta$ and the Hodge duality
$*$ operation are defined (see, e.g., [3]).

The off-shell nilpotent (anti-)BRST invariant version of the Lagrangian
density (2.11) for the 4D free 2-form Abelian gauge theory that
includes bosonic (anti-)ghost fields $(\bar\beta) \beta$
(with $\beta^2 \neq 0, \bar \beta^2 \neq 0$)
as well as the fermionic vector (anti-)ghost fields $(\bar C_\mu) C_\mu$
(with $C_\mu C_\nu + C_\nu C_\mu = 0, C_\mu \bar C_\nu + \bar C_\nu C_\mu
= 0, \bar C_\mu \bar C_\nu + \bar C_\nu \bar C_\mu = 0,
(C_\mu)^2 = 0, (\bar C_\mu)^2 = 0$ etc.) is [14]
$$
\begin{array}{lcl}
{\cal L}_{b1}^{(2)} &=& \frac{1}{12} H^{\mu\nu\kappa}
H_{\mu\nu\kappa} +  B^\mu \;(\partial^{\kappa} B_{\kappa\mu} -
\partial_\mu \phi_{1}) - \frac{1}{2} B^\mu B_\mu -
\partial_\mu \bar\beta \partial^\mu \beta \nonumber\\ &+&
(\partial_\mu \bar C_\nu - \partial_\nu \bar C_\mu)\;
(\partial^\mu C^\nu) +\; \rho\; (\partial \cdot C + \lambda) +\;
(\partial \cdot \bar C + \rho)\; \lambda,
\end{array} \eqno(3.7)
$$
where $B_\mu$ is the bosonic auxiliary field and $(\rho)\lambda$ are
the fermionic ($\rho^2 = \lambda^2 = 0, \rho \lambda + \lambda \rho = 0$)
scalar (anti-)ghost auxiliary fields. In fact, one can linearize the
quadratic kinetic energy term ($\frac{1}{12}
H^{\mu\nu\kappa} H_{\mu\nu\kappa}$) by introducing a couple of bosonic
auxiliary fields ${\cal B}_\mu$ and $\phi_{2}$ as [14,15]
$$
\begin{array}{lcl}
{\cal L}_{b2}^{(2)} &=& \frac{1}{2} {\cal B}^\mu {\cal B}_\mu -
{\cal B}^\mu\; (\frac{1}{2} \varepsilon_{\mu\nu\kappa\zeta}
\partial^\nu B^{\kappa\zeta} - \partial_\mu \phi_{2}) +
B^\mu \;(\partial^{\kappa} B_{\kappa\mu} - \partial_\mu \phi_{1})
- \frac{1}{2} B^\mu B_\mu \nonumber\\ &-& \partial_\mu \bar\beta
\partial^\mu \beta + (\partial_\mu \bar C_\nu - \partial_\nu \bar
C_\mu)\; (\partial^\mu C^\nu) + \rho\; (\partial \cdot C +
\lambda) + (\partial \cdot \bar C + \rho)\; \lambda.
\end{array} \eqno(3.8)
$$
The above Lagrangian density respects both the off-shell nilpotent
($s_{(D)B}^2 = 0$) (co-)BRST ($s_{(D)B}$) symmetry transformations as
listed below [14,15]
$$
\begin{array}{lcl}
s_{B} B_{\mu\nu} &=& (\partial_\mu C_\nu - \partial_\nu C_\mu), \qquad
s_{B} C_\mu = \partial_\mu \beta, \qquad s_{B} \bar C_\mu = B_\mu, \nonumber\\
s_{B} \phi_{1} &=& - \lambda,\; \qquad s_{B} \bar \beta = \rho, \;\qquad
\;\;\;s_{B} (\beta, {\cal B}_\mu, \lambda, \rho, B_\mu, \phi_{2},
H_{\mu\nu\kappa}) = 0,
\end{array} \eqno(3.9)
$$
$$
\begin{array}{lcl}
s_{D} B_{\mu\nu} &=& \varepsilon_{\mu\nu\kappa\zeta}
\partial^\kappa \bar C^\zeta, \quad s_{D} \bar C_\mu = - \partial_\mu
\bar \beta, \quad s_{D} C_\mu = {\cal B}_\mu, \quad s_{D} \beta = \lambda,
\nonumber\\
s_{D} \phi_{2} &=& - \rho, \qquad\;\;\;\;
s_{D} (\bar \beta, {\cal B}_\mu, \lambda, \rho, B_\mu, \phi_{1}, \partial^\mu
B_{\mu\kappa}) = 0.
\end{array} \eqno(3.10)
$$
It is interesting to point out that the off-shell nilpotent BRST symmetries
(3.9) are also respected by the Lagrangian density (3.7). If we substitute the
following equations of motion emerging from the Lagrangian density (3.8)
$$
\begin{array}{lcl}
B_\mu &=& (\partial^\nu B_{\nu\mu} - \partial_\mu \phi_{1}), \qquad\;\;\;
\rho = - \frac{1}{2}\; (\partial \cdot \bar C), \nonumber\\
{\cal B}_\mu &=& (\frac{1}{2} \varepsilon_{\mu\nu\kappa\zeta} \partial^\nu
B^{\kappa\zeta} - \partial_\mu \phi_{2}), \qquad
\lambda = - \frac{1}{2}\; (\partial \cdot C),
\end{array} \eqno(3.11)
$$
into (3.8) itself, we obtain the following Lagrangian  density
$$
\begin{array}{lcl}
{\cal L}_{b3}^{(2)} &=&
\frac{1}{2} (\partial^{\kappa} B_{\kappa\mu}
- \partial_\mu \phi_{1}) (\partial_\nu B^{\nu\mu} - \partial^\mu \phi_{1})
\nonumber\\
&-& \frac{1}{2} (\frac{1}{2} \varepsilon_{\mu\nu\kappa\zeta} \partial^\nu
B^{\kappa\zeta} - \partial_\mu \phi_{2}) (\frac{1}{2}
\varepsilon^{\mu\xi\sigma\eta} \partial_\xi B_{\sigma\eta}
- \partial^\mu \phi_{2}) \nonumber\\
&-& \partial_\mu \bar\beta \partial^\mu \beta
+ (\partial_\mu \bar C_\nu - \partial_\nu \bar C_\mu)\; (\partial^\mu C^\nu)
- \frac{1}{2}\; (\partial \cdot \bar C) (\partial \cdot C),
\end{array} \eqno(3.12)
$$
which turns out to respect the on-shell ($ \Box B_{\mu\nu} = \Box \phi_{1}
= \Box \phi_{2} = \Box \beta = \Box \bar \beta = 0, \Box C_\mu = \frac{3}{2}
\partial_\mu (\partial \cdot C), \Box \bar C_\mu = \frac{3}{2}
\partial_\mu (\partial \cdot \bar C)$)
nilpotent ($\tilde s_{(d)b}^2 = 0$) version of the (co-)BRST symmetry
transformations as expressed below
$$
\begin{array}{lcl}
\tilde s_{b} B_{\mu\nu} &=& (\partial_\mu C_\nu - \partial_\nu C_\mu), \qquad
\tilde s_{b} C_\mu = \partial_\mu \beta, \qquad
\tilde s_{b} \bar C_\mu = (\partial^\nu B_{\nu\mu} - \partial_\mu \phi_{1}),
 \nonumber\\
\tilde s_{b} \phi_{1} &=& + \frac{1}{2} (\partial \cdot C), \qquad
\tilde s_{b} \bar \beta = - \frac{1}{2} (\partial \cdot \bar C), \qquad
\tilde s_{b} (\beta, \phi_{2}, H_{\mu\nu\kappa}) = 0,
\end{array} \eqno(3.13)
$$
$$
\begin{array}{lcl}
\tilde s_{d} B_{\mu\nu} &=& \varepsilon_{\mu\nu\kappa\zeta}
\partial^\kappa \bar C^\zeta, \quad \tilde s_{d} \bar C_\mu = - \partial_\mu
\bar \beta, \quad \tilde s_{d} C_\mu =
(\frac{1}{2} \varepsilon_{\mu\nu\kappa\sigma} \partial^\nu B^{\kappa\sigma}
- \partial_\mu \phi_{2}),
\nonumber\\
\tilde s_{d} \phi_{2} &=&  + \frac{1}{2} (\partial \cdot \bar C),
\qquad\;\;\;\;
\tilde s_{d} \beta = - \frac{1}{2} (\partial \cdot C), \quad
\tilde s_{d} (\bar \beta, \phi_{1}, \partial^\mu B_{\mu\kappa}) = 0.
\end{array} \eqno(3.14)
$$
There are specific pertinent points in order here. First,
it will be noted that the Lagrangian density (3.12) is the generalized
version of (2.12) which respects both the on-shell nilpotent (co-)BRST
symmetries. Second, the generalization of the discrete symmetry transformations
(2.14) for the Lagrangian density (3.8) (that respects the off-shell nilpotent
(co-)BRST symmetry transformations (3.9) and (3.10)) is as follows [14,15]
$$
\begin{array}{lcl}
B_{\mu\nu} &\rightarrow& \mp \frac{i}{2} \varepsilon_{\mu\nu\kappa\xi}
B^{\kappa\xi}, \qquad \phi_{1} \rightarrow \pm i \phi_{2}, \qquad
\phi_{2} \rightarrow \mp i \phi_{1}, \nonumber\\
B_\mu &\rightarrow& \pm i {\cal B}_\mu, \quad
{\cal B}_\mu \rightarrow \mp i B_\mu, \quad \beta \rightarrow \mp i \bar \beta,
\quad \bar \beta \rightarrow \pm i \beta, \nonumber\\
C_\mu &\rightarrow& \pm i \bar C_\mu, \quad \bar C_\mu \rightarrow
\pm i C_\mu,
\quad \rho \rightarrow \pm i \lambda, \quad \lambda \rightarrow \pm i \rho.
\end{array}\eqno (3.15)
$$
The on-shell version of the above discrete symmetry transformations for the
Lagrangian density (3.12) can be obtained from the substitution of the
equations of motion (3.11) for the auxiliary fields into (3.15). Third, the
on-shell as well as off-shell version of the anti-BRST and anti-co-BRST
transformations can be obtained from (3.9), (3.10), (3.13) and (3.14)
by exploiting the discrete transformations (3.15) for the bosonic as well
as fermionic (anti-)ghost fields.
Finally, all the symmetry transformations of the theory can be generically
expressed in terms of the symmetry generators $Q_{r}$ as (see, e.g., [10,14])
$$
\begin{array}{lcl}
s_{r} \Phi = - i \; [ \Phi, Q_{r} ]_{\pm}, \qquad r = b, ab, d, ad, g, w,
\end{array} \eqno(3.16)
$$
where the subscripts $(+)-$ on the brackets correspond to (anti-)commutators
for the generic field $\Phi$ being (fermionic)bosonic in nature
and $g, w$ correspond to the
existence of a ghost symmetry and a bosonic symmetry. The above
generic expression is valid for the 2D as well as 4D theories for the off-shell
as well as on-shell nilpotent version of the (anti-) BRST and (anti-)co-BRST
symmetries together with other symmetries of the theories [10,14].\\

\noindent
{\bf 4 BRST cohomology: physical state condition}\\

\noindent
Here we shall recall some of the key
and pertinent points of earlier work (see, e.g., [9])
on the Hodge decomposition theorem for the 2D free Abelian gauge theory. To
this end in mind, we first express the normal mode expansion for the
basic fields ($A_\mu, C, \bar C)$ of the Lagrangian density (3.1) in the
(momentum) phase space as (see, e.g.,[9,37])
$$
\begin{array}{lcl}
A_\mu (x) &=& {\displaystyle \int \; \frac{d k} {(2\pi)^{1/2} (2 k^0)^{1/2}}}
\;\bigl [\; a_{\mu}^\dagger (k) e^{i k \cdot x} +
a_{\mu} (k) e^{- i k \cdot x}
\; \bigr ], \nonumber\\
C (x) &=& {\displaystyle \int \; \frac{d k} {(2\pi)^{1/2} (2 k^0)^{1/2}}}
\;\bigl [ \;c^\dagger (k) e^{i k \cdot x} + c (k) e^{- i k \cdot x}
\; \bigr ], \nonumber\\
\bar C (x) &=& {\displaystyle \int \; \frac{d k} {(2\pi)^{1/2} (2 k^0)^{1/2}}}
\; \bigl [\; \bar c^\dagger (k) e^{i k \cdot x} + \bar c (k) e^{- i k \cdot x}
\; \bigr ],
\end{array} \eqno(4.1)
$$
which corresponds to the equations of motion $\Box A_\mu = \Box C = \Box
\bar C = 0$ obeyed by the basic fields of the theory. Here $k_\mu$ are the
2D momenta with their components $ k_\mu = (k_{0}, k = k_{1})$ and
$k^2 = k_{0}^2 - k^2 = 0$ for the consistency with the above equations of
motion. All the dagger operators are the creation operators and
the non-dagger operators correspond to the
annihilation operators for the basic quanta of the fields. The on-shell
nilpotent version of the (co-)BRST symmetries (3.3) and (3.2) can be expressed,
due to (3.16), as [9,37]
$$
\begin{array}{lcl}
&& [ Q_{d}, a_{\mu}^\dagger (k) ] = - \varepsilon_{\mu\nu} k^\nu
\bar c^\dagger (k), \qquad
[ Q_{d}, a_{\mu} (k) ] = \varepsilon_{\mu\nu} k^\nu \bar c (k), \nonumber\\
&& \{ Q_{d}, c^\dagger (k) \} = - i \varepsilon^{\mu\nu}
k_\mu a_{\nu}^\dagger, \qquad
 \{ Q_{d}, c (k) \} =  + i \varepsilon^{\mu\nu} k_\mu a_\nu, \nonumber\\
&& \{ Q_{d}, \bar c^\dagger (k) \} =  0, \;\;\;\;\;\qquad\;\;\;\;\;
 \{ Q_{d}, \bar c (k) \} =  0,
\end{array} \eqno(4.2)
$$
$$
\begin{array}{lcl}
&& [ Q_{b}, a_{\mu}^\dagger (k) ] = k_\mu c^\dagger (k), \qquad
[ Q_{b}, a_{\mu} (k) ] = - k_\mu c (k), \nonumber\\
&& \{ Q_{b}, c^\dagger (k) \} = 0, \;\;\;\;\;\qquad \;\;\;\;\;
 \{ Q_{b}, c (k) \} = 0, \nonumber\\
&& \{ Q_{b}, \bar c^\dagger (k) \} =  i k^\mu a_{\mu}^\dagger (k), \qquad
 \{ Q_{b}, \bar c (k) \} =  - i k^\mu a_\mu (k).
\end{array} \eqno(4.3)
$$
Similar kinds of (anti-)commutation relations can be obtained with the
anti-BRST and anti-co-BRST generators but we do not require them for our
present analyses and
discussions. For aesthetic reasons, we can define the most symmetric
physical vacuum ($|vac>$) of the present theory as
$$
\begin{array}{lcl}
&&Q_{(a)b} \;| vac > = 0, \qquad Q_{(a)d}\; | vac > = 0,
\qquad Q_{w} \;|vac > = 0,
\nonumber\\
&& a_{\mu} (k)\; | vac > = 0, \qquad c (k)\; | vac > = 0, \qquad \;\;
\bar c (k)\; |vac > = 0.
\end{array}\eqno(4.4)
$$
In the above, it is clear that the physical vacuum is (anti-)BRST and
(anti-)co-BRST invariant which imply the invariance w.r.t. $Q_{w}$ as well.
It is lucid now that
a single photon state with polarization $e_\mu (k) $ and momenta $k_\mu$ can be
created from the physical vacuum by the application of a creation operator
$a_\mu^\dagger (k)$ as [37]
$$
\begin{array}{lcl}
e^\mu a_{\mu}^\dagger (k)\; | vac > \equiv | e, vac >, \qquad
k^\mu a_{\mu}^\dagger (k)\; | vac > \equiv | k, vac > = -  i \{ Q_{b}
\bar c^\dagger (k) \} \;|vac>,
\end{array} \eqno(4.5)
$$
where the latter state $ |k, vac>$ with momenta $k_\mu$ has been expressed by
exploiting the anti-commutator $\{ Q_{b}, \bar c^\dagger (k) \} = i k^\mu
a_\mu^\dagger (k)$ from (4.3). Exploiting the $U(1)$ gauge transformation
on the polarization vector $e_\mu (k) \rightarrow e_\mu^\prime (k) =
e_\mu (k) + i A k_\mu$, where $A$ is a complex number, it is straightforward
to check that
$$
\begin{array}{lcl}
| e + i\; A\; k, vac > = | e, vac > + \;
Q_{b}\; (A\; \bar c^\dagger (k)) | vac >, \qquad Q_{b}\; |vac> = 0.
\end{array} \eqno(4.6)
$$
Thus, we conclude that a gauge transformed state for an original single photon
state (i.e. $ e^\mu (k) a_\mu^\dagger (k) |vac>$
with the polarization vector $e_\mu (k)$) is equal to the sum of the original
state $| e, vac>$ plus a BRST exact state. In more sophisticated language,
the gauge transformed state and the original state belong to the same
cohomology class w.r.t. the
conserved and nilpotent BRST charge $Q_{b}$.
Similarly, the dual gauge transformation on the polarizations vector
(i.e. $ e_\mu (k) \rightarrow e_\mu^\prime (k) = e_\mu (k) +
i B \varepsilon_{\mu\nu}
k^\nu \equiv e_\mu (k) + i B \tilde k_\mu$ where $\tilde k_\mu =
\varepsilon_{\mu\nu} k^\nu$ and B is a complex number) will correspond to the
following expression
$$
\begin{array}{lcl}
| e + i\; B \tilde k, vac > = | e, vac > + \;Q_{d} \;(B\;
c^\dagger (k)) \;| vac >, \qquad Q_{d}\; |vac> = 0,
\end{array} \eqno(4.7)
$$
where we have used the anti-commutator $\{ Q_{d}, c^\dagger (k)\} =
- i \varepsilon^{\mu\nu} k_\mu a_\nu^\dagger$.
The above equation also implies that the dual-gauge transformed state is
equal to the sum of the original state and a BRST co-exact state.
With the four nilpotent and conserved charges $Q_{(a)b}, Q_{(a)d}$ and
a  bosonic conserved charge $Q_{w}$ in the theory, the most symmetric
physical state ($| phy >$) can be defined as
$$
\begin{array}{lcl}
Q_{(a)b}\; | phy > = 0, \qquad Q_{(a)d} \;| phy > = 0, \qquad
Q_{w} \;| phy > = 0.
\end{array} \eqno(4.8)
$$
Applying this physicality condition on the single photon state, we obtain the
following relationships by exploiting the commutators
$ [ Q_{b}, a_\mu^\dagger (k) ] = k_\mu c^\dagger (k), [Q_{d}, a_\mu^\dagger (k)
] = - \varepsilon_{\mu\nu} k^\nu \bar c^\dagger (k)$
$$
\begin{array}{lcl}
 Q_{b} \;| e + i\; A \;k,  vac> &=& Q_{b} \;| e, vac > \equiv (k \cdot e)\;
c^\dagger (k)\; | vac> = 0, \qquad
(Q_{b}^2 = 0),  \nonumber\\
 Q_{d} | e + i\; B \;\tilde k,  vac> &=& Q_{d} \;| e, vac > \equiv
(-\varepsilon^{\mu\nu} e_\mu k_\nu)\; \bar c^\dagger (k)\; |vac> = 0, \quad
(Q_{d}^2 = 0),
\end{array} \eqno(4.9)
$$
which imply the {\it transversality} (i.e. $ k \cdot e = 0$) and an
extra condition ($ \varepsilon^{\mu\nu} e_\mu k_\nu = 0$)
(that turns out to be useful in the proof for the topological nature of
the 2D free Abelian gauge theory [9,10]) on the 2D photon because of the fact
that $ c^\dagger (k) | vac > \neq 0, \bar c^\dagger (k) |vac> \neq 0$. It is
also obvious from the above discussion that for a single photon, {\it not}
satisfying the above transversality and an extra condition
illustrated in (4.9), the single
(anti-)ghost state(s) ($\bar c^\dagger (k) |vac>, c^\dagger (k) |vac>$)
created by the operators $\bar c^\dagger (k)$ and $c^\dagger (k)$ would
turn out to be BRST (co-)exact states. This explains the {\it no-(anti-)ghost}
theorem in the context of the BRST cohomology. Physically, it amounts to
the well-known fact (see, e.g., [42]) that the
contributions coming from the longitudinal and
scalar degrees of freedom of
the photons, at any arbitrary order of the perturbation
theory calculations, are cancelled by the presence of (anti-)ghost fields.
Ultimately, the physicality
criteria $Q_{b} | e, vac> = 0, Q_{d} |e, vac> = 0, Q_{w} |e, vac> = 0$
on a single photon state implies the transversality and masslessness
of the photon because of the following {\it mutually consistent} relationships
that emerge from the above condition with conserved
and local charges (see, e.g., [9] for more detail)
$$
\begin{array}{lcl}
&&Q_{b} \;| e, vac> = 0 \;\rightarrow \;(k \cdot e) = 0, \nonumber\\
&&Q_{d} \;|e, vac> = 0 \;\rightarrow \;
(\varepsilon_{\mu\nu} e^\mu k^\nu) = 0,
\nonumber\\ && Q_{w} \;|e, vac> = 0,
\;\rightarrow \;k^2 = 0.
\end{array} \eqno(4.10)
$$
A thorough and complete
 discussion on this result and its implication to the topological nature
of the theory (where there are no propagating degrees of freedom for the
$U(1)$ gauge field $A_\mu$) can be found in our earlier works
(see, e.g., [9,10]). In the
language of the Hodge decomposition theorem, it can be seen that the
masslessness condition ($k^2 = 0$) coming from $Q_{w} |e, vac> = 0$
(which is the analogue of the action of the Laplacian operator on the harmonic
state) has the solutions $ k \cdot e = 0$ and
$ \varepsilon_{\mu\nu} e^\mu k^\nu = 0$ that are given by the top two
equations (i.e., the analogues of the operation of the
(co-)exterior derivatives).

With our understanding of the 2D free Abelian gauge theory as the background,
we shall dwell a bit on the BRST cohomology connected with the 4D free Abelian
2-form gauge theory. For this purpose, first, we begin with
the definition of the physical state $|phys>$ and
the physical vacuum ($ |vacm>$) of the 4D theory as
$$
\begin{array}{lcl}
 Q_{(A)B} \;|phys> =  0, \qquad Q_{(A)D} \;|phys> = 0,
\qquad Q_{W} \;|phys> = 0,
\end{array} \eqno(4.11)
$$
$$
\begin{array}{lcl}
&& Q_{(A)B} |vacm> =  0, \qquad Q_{(A)D} |vacm> = 0, \qquad Q_{W} |vacm> = 0,
\nonumber\\
&& b_{\mu\nu} (k)\; |vacm> = 0, \qquad c_\mu (k)\; |vacm> = 0, \qquad
\bar c_\mu (k)\; |vacm> =  0, \nonumber\\
&& \bar b (k) \;|vacm> = 0, \quad f_{1} (k) \;|vacm> = 0,                       \quad f_{2} (k) \; |vacm> = 0,
\quad  b (k) |vacm> = 0,
\end{array} \eqno(4.12))
$$
where {\it the on-shell nilpotent} (anti-)BRST charges $Q_{(A)B}$ and
(anti-)co-BRST charges $Q_{(A)D}$ are the generators of the
corresponding transformations
(3.13) and (3.14) for the Lagrangian density (3.12). Here the bosonic charge
$Q_{W} = \{ Q_{B}, Q_{D} \} = \{Q_{AB}, Q_{AD} \}$ and the rest of the
annihilation operators in the above are from
the normal mode expansion of the basic fields of the theory that are
present in the Lagrangian density (3.12). These expansions are as follows
$$
\begin{array}{lcl}
B_{\mu\nu} (x) &=& {\displaystyle \int \;
\frac{d^3 k} {(2\pi)^{3/2} (2 k^0)^{3/2}}}
\;\bigl [\; b_{\mu\nu}^\dagger (k) e^{i k \cdot x} +
b_{\mu\nu} (k) e^{- i k \cdot x}
\; \bigr ], \nonumber\\
C_\mu (x) &=& {\displaystyle \int \;
\frac{d^3 k} {(2\pi)^{3/2} (2 k^0)^{3/2}}}
\;\bigl [ \;c_\mu^\dagger (k) e^{i k \cdot x} + c_\mu (k) e^{- i k \cdot x}
\; \bigr ], \nonumber\\
\bar C_\mu (x) &=& {\displaystyle \int \;
\frac{d^3 k} {(2\pi)^{3/2} (2 k^0)^{3/2}}}
\; \bigl [\; \bar c_\mu^\dagger (k) e^{i k \cdot x}
+ \bar c_\mu (k) e^{- i k \cdot x}
\; \bigr ], \nonumber\\
\bar \beta (x) &=& {\displaystyle \int \;
\frac{d^3 k} {(2\pi)^{3/2} (2 k^0)^{3/2}}}
\; \bigl [\; \bar b^\dagger (k) e^{i k \cdot x}
+ \bar b (k) e^{- i k \cdot x}
\; \bigr ], \nonumber\\
\beta (x) &=& {\displaystyle \int \;
\frac{d^3 k} {(2\pi)^{3/2} (2 k^0)^{3/2}}}
\; \bigl [\; b^\dagger (k) e^{i k \cdot x}
+ b (k) e^{- i k \cdot x}
\; \bigr ], \nonumber\\
\phi_{1} (x) &=& {\displaystyle \int \;
\frac{d^3 k} {(2\pi)^{3/2} (2 k^0)^{3/2}}}
\; \bigl [\; f_{1}^\dagger (k) e^{i k \cdot x}
+ f_{1} (k) e^{- i k \cdot x}
\; \bigr ], \nonumber\\
\phi_{2} (x) &=& {\displaystyle \int \;
\frac{d^3 k} {(2\pi)^{3/2} (2 k^0)^{3/2}}}
\; \bigl [\; f_{2}^\dagger (k) e^{i k \cdot x}
+ f_{2}(k) e^{- i k \cdot x}
\; \bigr ], \nonumber\\
\end{array} \eqno(4.13)
$$
where the four momenta $k_\mu$ have the components $k_\mu = (k_{0},
k_{i})$ with $ i = 1, 2, 3 $ and the volume
element $ d^3 k = d k_{1} d k_{2} d k_{3}$. All the individual
dagger operators are the creation operator for a single quantum (particle)
of the corresponding
field (and the non-dagger operators are the annihilation operators). Exploiting
the general definition of (3.16), we obtain the (anti-)commutators with
the BRST charge $Q_{B}$ as
$$
\begin{array}{lcl}
&& [ Q_{B}, b_{\mu\nu}^\dagger (k) ] = (k_\mu c_\nu^\dagger (k)
- k_\nu c_\mu^\dagger (k)), \qquad
[ Q_{B}, b_{\mu\nu} (k) ] = - (k_\mu c_\nu (k) - k_\nu c_\mu (k)), \nonumber\\
&& [ Q_{B}, b^\dagger (k) ]  = 0, \;\;\;\;\;\qquad \;\;\;\;\;
 [ Q_{B}, b (k) ] = 0, \nonumber\\
&& \{ Q_{B}, \bar c_\mu^\dagger (k) \} =  + k^\nu
b_{\mu\nu}^\dagger (k) + k_\mu f_{1}^\dagger (k), \qquad
 \{ Q_{B}, \bar c_\mu (k) \} = - k^\nu b_{\mu\nu} (k) - k_\mu f_{1} (k),
\nonumber\\
&& \{ Q_{B},  c_\mu^\dagger (k) \} =  - k_\mu b^\dagger (k), \qquad
 \{ Q_{B}, c_\mu (k) \} =  + k_\mu b (k),
\nonumber\\
&& [ Q_{B}, f_{2}^\dagger (k) ] = 0, \;\;\;\;\;\qquad \;\;\;\;\;
 [ Q_{B}, f_{2} (k) ] = 0, \nonumber\\
&& [ Q_{B}, f_{1}^\dagger (k) ] = \frac{1}{2} k^\mu c^\dagger_\mu (k),
\;\;\;\;\;\qquad \;\;\;\;\;
 [ Q_{B}, f_{1} (k) ] = - \frac{1}{2} k^\mu c_\mu (k), \nonumber\\
&& [ Q_{B}, \bar b^\dagger (k) ] = - \frac{1}{2} k^\mu \bar
c^\dagger_\mu (k), \qquad
 [ Q_{B},  \bar b (k) ] = + \frac{1}{2} k^\mu \bar c_\mu (k),
\end{array} \eqno(4.14)
$$
and the corresponding (anti-)commutators with the co-BRST charge $Q_{D}$ are
$$
\begin{array}{lcl}
&& [ Q_{D}, b_{\mu\nu}^\dagger (k) ] =
\varepsilon_{\mu\nu\eta\zeta} k^\eta (\bar c^\zeta)^\dagger (k), \qquad
[ Q_{D}, b_{\mu\nu} (k) ] = - \varepsilon_{\mu\nu\eta\zeta} k^\eta
\bar c^\zeta (k), \nonumber\\
&& [ Q_{D}, \bar b^\dagger (k) ]  = 0, \;\;\;\;\;\qquad \;\;\;\;\;
 [ Q_{D}, \bar b (k) ] = 0, \nonumber\\
&& \{ Q_{D}, \bar c_\mu^\dagger (k) \} =  k_\mu \bar b^\dagger (k), \qquad
 \{ Q_{D}, \bar c_\mu (k) \} = - k_\mu \bar b (k),
\nonumber\\
&& \{ Q_{D},  c_\mu^\dagger (k) \} =
k_\mu f_{2}^\dagger (k)
- \frac{1}{2} \varepsilon_{\mu\nu\kappa\sigma} k^\nu
(b^{\kappa\sigma})^\dagger (k), \nonumber\\
&& \{ Q_{B}, c_\mu (k) \} =
- k_\mu f_{2} (k)
+ \frac{1}{2} \varepsilon_{\mu\nu\kappa\sigma} k^\nu
b^{\kappa\sigma} (k),
\nonumber\\
&& [ Q_{D}, f_{1}^\dagger (k) ] = 0, \;\;\;\;\;\qquad \;\;\;\;\;
 [ Q_{D}, f_{1} (k) ] = 0, \nonumber\\
&& [ Q_{D}, f_{2}^\dagger (k) ] = \frac{1}{2} k^\mu \bar c^\dagger_\mu (k),
\;\;\;\;\;\qquad \;\;\;\;\;
 [ Q_{D}, f_{2} (k) ] = - \frac{1}{2} k^\mu \bar c_\mu (k), \nonumber\\
&& [ Q_{D}, b^\dagger (k) ] = - \frac{1}{2} k^\mu c^\dagger_\mu (k),
\qquad
 [ Q_{D},  \bar b (k) ] = + \frac{1}{2} k^\mu c_\mu (k).
\end{array} \eqno(4.15)
$$
A few comments are in order now. First of all, it can be seen that the
expansions for the vector anti-commuting (anti-)ghost fields
$(\bar C_\mu ) C_\mu $ in (4.13) are not consistent with the equations
of motion $\Box C_\mu = \frac{3}{2} \;\partial_\mu (\partial \cdot C),\;
\Box \bar C_\mu = \frac{3}{2} \;\partial_\mu (\partial \cdot \bar C)$
 unless we choose a gauge such that $(\partial \cdot C) =
(\partial \cdot \bar C) = 0$ thereby implying
 $\Box C_\mu = \Box \bar C_\mu = 0$.
Second, the above choice will imply $ k^2 = 0, \;k\cdot c^\dagger =
k \cdot c = 0, \;k \cdot \bar c^\dagger = k \cdot \bar c = 0$ in the
(momentum) phase space. Third, it
is evident that the last four commutators of (4.14) and (4.15) will turn
out to be zero for the above gauge choice. We shall come to these points
in a more detailed fashion in the next section where we will discuss the
connection between little group and cohomology.

Let us now concentrate on a single particle quantum state for the
2-form field with polarization
tensor $e_{\mu\nu} (k) = - e_{\nu\mu} (k)$. This can be created from the vacuum
state by application of the creation operator $b_{\mu\nu}^\dagger (k)$ as
$$
\begin{array}{lcl}
e^{\mu\nu} (k) b_{\mu\nu}^\dagger (k) |vacm> \equiv | \tilde e, vacm>.
\end{array} \eqno(4.16)
$$
The physicality criteria on this single particle
quantum state for the 2-form field leads to
$$
\begin{array}{lcl}
Q_{B} \; |\tilde e, vacm> = 0 &\rightarrow&
[ Q_{B}, e^{\mu\nu} (k) b_{\mu\nu}^\dagger (k) ]\; |vacm> = 0, \nonumber\\
Q_{D} \; |\tilde e, vacm> = 0 &\rightarrow&
[ Q_{D}, e^{\mu\nu} (k) b_{\mu\nu}^\dagger (k) ]\; |vacm> = 0, \nonumber\\
Q_{W} \; |\tilde e, vacm> = 0 &\rightarrow&
[ Q_{W}, e^{\mu\nu} (k) b_{\mu\nu}^\dagger (k) ]\; |vacm> = 0.
\end{array} \eqno(4.17)
$$
Exploiting the following commutation relations
$$
\begin{array}{lcl}
&& [ Q_{B}, b_{\mu\nu}^\dagger (k) ] = (k_\mu c_\nu^\dagger (k) - k_\nu
c_\mu^\dagger (k)), \qquad
[ Q_{D}, b_{\mu\nu}^\dagger (k) ] = \varepsilon_{\mu\nu\eta\xi}
k^\eta (\bar c^\xi)^\dagger (k), \nonumber\\
&& [ Q_{W}, b_{\mu\nu}^\dagger (k) ] = \frac{i}{2}\; [\;
\varepsilon_{\mu\nu\zeta\eta} k_\sigma
- \varepsilon_{\mu\nu\zeta\sigma} k_\eta
+ \varepsilon_{\nu\zeta\sigma\eta} k_\mu
- \varepsilon_{\mu\zeta\sigma\eta} k_\nu\; ]\; k^\zeta
(b^{\sigma\eta})^\dagger (k),
\end{array} \eqno(4.18)
$$
where the top two equations have been taken from (4.14) and (4.15)
and the last commutation relation has been extracted from
the following bosonic symmetry transformations $s_{W}
= \{s_{B}, s_{D} \}$ for the Lagrangian density (3.12)
$$
\begin{array}{lcl}
&& s_{W} B_{\mu\nu} = \varepsilon_{\mu\nu\eta\zeta} \partial^\eta
(\partial_\sigma B^{\sigma\zeta})
+ \frac{1}{2} \varepsilon_{\nu\zeta\sigma\eta} \partial_\mu (\partial^\zeta
B^{\sigma\eta})
- \frac{1}{2} \varepsilon_{\mu\zeta\sigma\eta} \partial_\nu (\partial^\zeta
B^{\sigma\eta}), \nonumber\\
&& s_{W} C_\mu (x) = - \frac{1}{2} \partial_\mu (\partial \cdot C), \quad
s_{W} \bar C_\mu (x) = + \frac{1}{2} \partial_\mu (\partial \cdot \bar C),
\quad s_{W} (\phi_{1}, \phi_{2}, \beta, \bar \beta) = 0.
\end{array} \eqno(4.19)
$$
These can be recast in terms of the commutation relations between $Q_{W}$
and the creation and annihilation operators for the fields as
$$
\begin{array}{lcl}
&& [ Q_{W}, b_{\mu\nu}^\dagger (k) ] = \frac{i}{2}\; [ \;
\varepsilon_{\mu\nu\zeta\eta} k_\sigma
- \varepsilon_{\mu\nu\zeta\sigma} k_\eta
+ \varepsilon_{\nu\zeta\sigma\eta} k_\mu
- \varepsilon_{\mu\zeta\sigma\eta} k_\nu \;]\; k^\zeta\;
(b^{\sigma\eta})^\dagger (k),
\nonumber\\
&& [ Q_{W}, b_{\mu\nu} (k) ] = \frac{i}{2}\; [\;
\varepsilon_{\mu\nu\zeta\eta} k_\sigma
- \varepsilon_{\mu\nu\zeta\sigma} k_\eta
+ \varepsilon_{\nu\zeta\sigma\eta} k_\mu
- \varepsilon_{\mu\zeta\sigma\eta} k_\nu \;]\; k^\zeta\; b^{\sigma\eta} (k),
\nonumber\\
&& [ Q_{W}, c_\mu^\dagger (k) ] = - \frac{i}{2} k_\mu \;
(k^\nu c_{\nu}^\dagger (k) ), \qquad
[ Q_{W}, c_\mu (k) ] = + \frac{i}{2} k_\mu \;
(k^\nu c_{\nu} (k)), \nonumber\\
&& [ Q_{W}, \bar c_\mu^\dagger (k) ] = + \frac{i}{2} k_\mu \;
(k^\nu \bar c_{\nu}^\dagger (k) ), \qquad
[ Q_{W}, \bar c_\mu (k) ] = - \frac{i}{2} k_\mu \;
(k^\nu \bar c_{\nu} (k)).
\end{array} \eqno(4.20)
$$
It is evident that the physicality conditions in (4.17) (which are the
analogues of (4.10) for the 2D free Abelian theory) lead to the
following restrictions
$$
\begin{array}{lcl}
&&e^{\mu\nu} (k)\; [\;k_\mu c_\nu^\dagger (k)
- k_\nu c_\mu^\dagger (k)\;]\; |vacm> = 0
\rightarrow - 2\; [e^{\mu\nu} (k) k_\nu)]\; c_{\mu}^\dagger (k) \;|vacm> = 0,
\nonumber\\
&&e^{\mu\nu} (k) \; [\;\varepsilon_{\mu\nu\eta\zeta} k^\eta \;
(\bar c^\zeta)^\dagger (k)\;] \; |vacm> = 0
\rightarrow [\;\varepsilon_{\mu\nu\eta\zeta} e^{\mu\nu} (k) k^\eta \;] \;
(\bar c^{\zeta})^\dagger (k) \;|vacm> = 0, \nonumber\\
&&\frac{i}{2}\; e^{\mu\nu} (k) \;[ \;
\varepsilon_{\mu\nu\zeta\eta} k_\sigma
- \varepsilon_{\mu\nu\zeta\sigma} k_\eta
+ \varepsilon_{\nu\zeta\sigma\eta} k_\mu
- \varepsilon_{\mu\zeta\sigma\eta} k_\nu \;]\; k^\zeta \;
(b^{\sigma\eta})^\dagger (k) \; |vacm> = 0 \nonumber\\
&& \rightarrow i [\;\varepsilon_{\mu\nu\zeta\eta} k_\sigma
- \varepsilon_{\mu\zeta\sigma\eta} k_\nu \;]\;e^{\mu\nu} (k) \;k^\zeta \;
(b^{\sigma\eta})^\dagger (k) \; |vacm> = 0.
\end{array} \eqno(4.21)
$$
The top equation establishes the transversality condition for the
2-form gauge field because $k_\mu \varepsilon^{\mu\nu} \equiv
- \varepsilon^{\mu\nu} k_\nu = 0$ shows that the polarization tensor
and momenta are orthogonal to one-another. We draw this conclusion
because $ c_\mu^\dagger (k) \; |vacm> \neq 0$ and it actually creates a quantum
of the ghost field $C_\mu (x)$. For the transversality {\it not} to
be satisfied, the above relation demonstrates the famous {\it no-ghost theorem}
in the language of the BRST cohomology in the sense  that the state
$ c_\mu^\dagger (k) \;|vacm>$ is a BRST exact state and, hence,  a
cohomologically trivial state. The next condition due to the physicality
criteria
(w.r.t. the conserved and nilpotent co-BRST charge $Q_{D} |phys> = 0$) finally
implies the dual-transversality condition:
 $\varepsilon_{\mu\nu\kappa\zeta} e^{\nu\kappa} (k) k^\zeta = 0$.
This relation is, in some sense, an extension of the the second
equation of (4.10) (valid for the 2D free Abelian gauge theory) to the
4D free Abelian 2-form gauge theory. This also demonstrates the fact that
if $\varepsilon_{\mu\nu\kappa\zeta} e^{\nu\kappa} (k) k^\zeta \neq 0$,
the one-particle vector anti-ghost state $\bar c_\mu^\dagger (k)\; |vacm>$
is a BRST co-exact state and, therefore, a trivial state as far as the
BRST cohomology, HDT and choice of the physical state to be the harmonic state
are concerned. This establishes the no-anti-ghost theorem (without
resorting to the implementation of the anti-BRST charge ($Q_{AB}$) in the
physicality criteria which also leads to the same conclusion). The last
condition of (4.21), with  the help of the transversality condition
$- k_\mu e^{\mu\nu} (k) = e^{\mu\nu} (k) k_\nu = 0$, leads to the masslessness
(i.e. $\Box B_{\mu\nu} (x) = 0 \rightarrow k^2 = k_{0}^2 - k_{i}^2 = 0$)
condition for the 2-form $B_{\mu\nu}$ field when we expand the whole
equation in terms of the physical components of $k^\mu$ and $e^{\mu\nu} (k)$.
Thus, ultimately, the criteria in (4.21) physically imply the
{\it transversality} and {\it masslessness} of the free Abelian 2-form
gauge field.\\

\noindent
{\bf 5 Wigner's little group: (dual-)gauge transformations and cohomology}\\

\noindent
We begin with the most general form of the Wigner's little
group matrix $\{ W^\mu_\nu (\theta, u, v) \}$ for a massless
particle moving along the $z$-direction of the 4D spacetime manifold
as [30-35]
$$
\begin{array}{lcl}
\{ W (\theta, u, v) \} =
\left ( \begin{array}{cccc}
\bigl (1 + {\displaystyle \frac{u^2 + v^2} {2}} \bigr ) &
\bigl (u cos \theta -  v sin \theta \bigr )
& \bigl (u sin \theta + v cos \theta \bigr ) & -
\bigl ({\displaystyle \frac{u^2 + v^2}{2}} \bigr )\\
u & cos \theta & sin \theta  & - u \\
v & - sin \theta & cos \theta & - v\\
\bigl ({\displaystyle \frac{u^2 + v^2}{2}} \bigr ) &
\bigl (u cos \theta - v sin \theta \bigr )
&\bigl  (u sin \theta + v cos \theta \bigr ) &
\bigl (1 - {\displaystyle \frac{u^2 + v^2} {2}} \bigr ) \\
\end{array} \right ),
\end{array} \eqno(5.1)
$$
where $\theta$ is the rotational parameter and $u, v$ are the translational
parameters defining $T(2)$ in the $xy$ plane. By definition, this matrix
preserves the four momentum $k^\mu = (\omega, 0, 0, \omega)^T$ of
a massless ($k^2 = 0$) particle with energy $\omega$ and it can be
factorized elegantly as
$$
\begin{array}{lcl}
(k^\mu) \rightarrow (k^\mu)^\prime
= W^\mu_\nu\; (k^\nu) = (k^\mu), \; \qquad
W (\theta, u. v) = R (\theta)\; W (0, u, v).
\end{array} \eqno(5.2)
$$
The matrix $R(\theta)$ in the above represents the rotation about the z-axis
$$
\begin{array}{lcl}
R (\theta) =
\left ( \begin{array}{cccc}
1 & 0 & 0 & 0 \\
0 & cos \theta & sin \theta & 0 \\
0 & - sin \theta  & cos \theta & 0 \\
0 & 0 & 0 & 1 \\
\end{array} \right ),
\end{array} \eqno(5.3)
$$
and the matrix $ \{ W(0,u,v) \}$ is found to be isomorphic
to the two parameter translation group $T(2)$ (i.e. $T(2) \sim W (0,u,v)$)
in the two-dimensional Euclidean plane
($xy)$ which is a plane perpendicular to the propagation of the light-like
(massless) particle along the z-direction. {\it It is crystal clear that there
exists no such kind of plane for the discussion of any arbitrary
gauge theory in $(1 + 1)$-dimension (2D) of spacetime.}

Now let us concentrate on the gauge transformation (2.6) on the 2-form free
Abelian basic gauge field $B_{\mu\nu}$ (i.e. $ B_{\mu\nu} \rightarrow
B_{\mu\nu}^{(g)} = B_{\mu\nu} + (\partial_\mu \alpha_\nu
- \partial_\nu \alpha_\mu)$). This transformation, in the language of the
anti-symmetric ($ e_{\mu\nu} (k) = - e_{\nu\mu} (k)$) polarization tensor
can be expressed, in its contravariant form, as [36]
$$
\begin{array}{lcl}
e^{\mu\nu} (k) \rightarrow (e^{\mu\nu})^{(g)} (k) = e^{\mu\nu} (k)
+ i \;[k^\mu \alpha^\nu (k) - k^\nu \alpha^\mu (k)].
\end{array} \eqno(5.4)
$$
For the massless ($ k^2 = 0$) 2-form
$B_{\mu\nu} (x)$ field, it can be seen,
from the physicality condition (4.21) (with the BRST charge $Q_{B}$) that
the momenta and the polarization tensor are orthogonal to one-another:
$k_\mu e^{\mu\nu} (k) = - e^{\mu\nu} (k) k_\nu = 0$.
This orthogonality condition can also be obtained from the equation of motion
 for the massless (i.e. $ k^2 = 0$) 2-form gauge field $B_{\mu\nu} (x)$
when the Lagrangian density contains only the kinetic energy term.
One has to put masslessness ($k^2 = 0$) condition, however, by hand in
the latter case [36]. This
transversality condition reduces the six independent components of the
anti-symmetric 4D tensor $e^{\mu\nu} (k)$ to three only (cf. (5.7) below). For
this to be seen explicitly, we have the following matrix product
$$
\begin{array}{lcl}
k_\mu e^{\mu\nu} = 0 \Rightarrow
(\omega, 0, 0, - \omega)
\left ( \begin{array}{cccc}
0 & e^{01} & e^{02} & e^{03} \\
- e^{01} & 0 & e^{12} & e^{13} \\
- e^{02} & - e^{12} & 0 & e^{23}\\
- e^{03} & -e^{13} & - e^{23} & 0\\
\end{array} \right ) = 0,
\end{array} \eqno(5.5)
$$
where the covariant version of momentum 4-vector is taken to be
$k_\mu = (\omega, 0, 0, - \omega)^T$ for our choice of the contravariant
momentum vector $k^\mu = (\omega, 0, 0, \omega)^T$. The above conditions
yield the following relationships among the components of the polarization
matrix $\{ e^{\mu\nu} (k) \}$
$$
\begin{array}{lcl}
e^{03} = 0,\; \qquad e^{01} + e^{13} = 0, \;\qquad e^{02} + e^{23} = 0.
\end{array} \eqno(5.6)
$$
Thus, the above transversality
condition (5.5) reduces the polarization tensor to
$$
\begin{array}{lcl}
\{ e^{\mu\nu} (k) \}_{(r)} =
\left ( \begin{array}{cccc}
0 & e^{01} & e^{02} & 0 \\
- e^{01} & 0 & e^{12} & - e^{01} \\
- e^{02} & - e^{12} & 0 & - e^{02}\\
0 & e^{01} & e^{02} & 0\\
\end{array} \right ).
\end{array} \eqno(5.7)
$$
Yet another reduction of the anti-symmetric tensor $\{ e^{\mu\nu} (k) \}$
can be achieved by the choice of the infinitesimal gauge parameters
$\alpha$'s of the transformation (5.4) for our choice of the 4-momenta
$k^\mu = (\omega, 0, 0, \omega)^T$. For instance, it can be checked that,
for the following choice of the local gauge parameters
$$
\begin{array}{lcl}
\alpha^{1} (k) = {\displaystyle \frac{i}{\omega}} \;e^{01} (k),\; \qquad
\alpha^{2}  (k) = {\displaystyle \frac{i}{\omega}} \;e^{02} (k),
\end{array} \eqno(5.8)
$$
one can gauge away the components $e^{01} (k)$ and $e^{02} (k)$
(i.e. $(e^{01}) (k) \rightarrow (e^{01} (k))^{(g)} = 0,\;
(e^{02}) (k) \rightarrow (e^{02} (k))^{(g)} = 0 $). Even through
$e^{03} (k) = 0$ due to (5.6), in the general gauge transformation (5.4)
 one has to choose $\alpha^{0} (k) = \alpha^{3} (k)$ for the identity
(i.e. $0 = 0$) corresponding
to the gauge transformation on $e^{03} (k)$, to be satisfied.
Thus, ultimately, we are left with the anti-symmetric
second-rank matrix $ \{ e^{\mu\nu} (k) \}$ with only one degree of freedom
as
$$
\begin{array}{lcl}
\{ e^{\mu\nu} (k) \}_{(R)} = \;e^{12} (k)\;
\left ( \begin{array}{cccc}
0 & 0 & 0 & 0 \\
0 & 0 & 1 & 0 \\
0 & - 1 & 0 & 0\\
0 & 0 & 0 & 0\\
\end{array} \right ).
\end{array} \eqno(5.9)
$$
At this juncture, we can exploit the crucial role played by the translation
subgroup $T(2) \sim W (0,u,v)$ of the Wigner's little group in generating
the gauge transformation for the polarization tensor as follows
$$
\begin{array}{lcl}
&& e^{\mu\nu} (k) \rightarrow (e^{\mu\nu})^\prime (k)
= (W)^\mu_\sigma \; (e^{\sigma\kappa} (k)) \; (W)_\kappa^\nu
\equiv W (0, u, v) \cdot e (k) \cdot W (0, u, v)^T,
\nonumber\\
&& e^{\mu\nu} (k) \rightarrow (e^{\mu\nu})^\prime (k) = e^{12} (k)\;
\left ( \begin{array}{cccc}
0 & - v & u & 0 \\
v & 0 & 1 & v \\
- u & -1 & 0 & - u\\
0 & - v & u & 0\\
\end{array} \right ),
\end{array} \eqno(5.10)
$$
where the matrix product has been denoted by the dot ($\cdot$) product. The
above transformation can be re-expressed as
$$
\begin{array}{lcl}
 \{ e^{\mu\nu} (k) \} \rightarrow \{ (e^{\mu\nu})^\prime (k) \} =
\{ e^{\mu\nu} (k) \}_{(R)}
+  e^{12} (k)\;
\left ( \begin{array}{cccc}
0 & - v & u & 0 \\
v & 0 & 0 & v \\
- u & 0 & 0 & - u\\
0 & - v & u & 0\\
\end{array} \right ),
\end{array} \eqno(5.11)
$$
where the untransformed matrix $\{ e^{\mu\nu} (k)\}_{(R)}$ is given in (5.9).
This is obviously a gauge transformation corresponding to (5.4) for the
following relationships among the parameters of the gauge symmetry group
and the translation $T(2)$ subgroup of the Wigner's little group
$$
\begin{array}{lcl}
\alpha^{1} (k) = {\displaystyle \frac{i v}{\omega}} \;e^{12} (k),\; \qquad
\alpha^{2} (k) = - {\displaystyle \frac{i u}{\omega}} \;e^{12} (k),\; \qquad
\alpha^{3} = \alpha^{0}.
\end{array} \eqno(5.12)
$$
It will be noted that the last condition comes from the consideration
of the gauge transformation (5.4) on the component $e^{03} (k)$
(or $ e^{30} (k)$). The other two conditions emerge from the gauge
transformations on $e^{01} (k), e^{02} (k),
e^{13} (k), e^{23} (k)....$ etc. This establishes the connection between
translation subgroup $T(2) \sim W (0,u,v)$ of the Wigner's little group and
the gauge transformation defined in (2.6).

Now we dwell a bit on the origin of the dual-gauge transformation
($B_{\mu\nu} (x) \rightarrow B_{\mu\nu}^{(dg)} (x) = B_{\mu\nu} (x)
+ \varepsilon_{\mu\nu\kappa\xi} \partial^\kappa \Sigma^\xi (x)$) defined
in (2.6) in the framework of Wigner's little group.
It is clear that this transformation in the momentum space can be expressed,
following [36], in terms of the transformation on the contravariant
polarization anti-symmetric tensor $e^{\mu\nu} (k)$ as
$$
\begin{array}{lcl}
(e^{\mu\nu}) (k) \rightarrow (e^{\mu\nu})^{(dg)} (k)
= (e^{\mu\nu}) (k) + i \varepsilon^{\mu\nu\eta\xi}
k_\eta \Sigma_\xi (k).
\end{array} \eqno(5.13)
$$
We would like to answer the question: in the framework of the translation
subgroup $T(2)$ of the Wigner's little group, where is the existence of the
dual-gauge transformation of (2.6) which has been expressed as (5.13)? It
is evident that the argument, based on the {\it transversality} of the 2-form
gauge field emerging from the physicality condition with the BRST
charge $Q_{B}$, goes along the same lines till equation (5.7) where
we have obtained the reduced polarization tensor $\{ e^{\mu\nu} (k) \}_{(r)}$.
It can be checked, using (5.13) and $k_\mu = (\omega, 0, 0, -\omega)^T$,
that for the following choice of the infinitesimal dual-gauge
parameters $\Sigma$'s
$$
\begin{array}{lcl}
\Sigma_{1} (k) = {\displaystyle \frac{i}{\omega} }\; e^{02} (k),\; \qquad
\Sigma_{2} (k) = - {\displaystyle \frac{i}{\omega} }\; e^{01} (k),
\end{array} \eqno(5.14)
$$
one can gauge away the components $e^{01} (k)$ and $e^{02} (k)$ from the matrix
$\{ e^{\mu\nu} (k) \}_{(r)}$
(i.e. $(e^{01} (k) \rightarrow (e^{01})^{(dg)} (k) = 0,
\; e^{02} (k) \rightarrow (e^{02})^{(dg)} (k) = 0 $).
We note here that there is a kind of ``duality''
between (5.8) and (5.14) in the sense that
$$
\begin{array}{lcl}
\alpha^{1} (k) \leftrightarrow - \Sigma_{2} (k),\; \qquad
\alpha^{2} (k) \leftrightarrow \Sigma_{1} (k).
\end{array}\eqno(5.15)
$$
The relationship in (5.14) finally allows us to obtain
the reduced matrix $ \{ e^{\mu\nu} (k) \}_{(R)}$ of (5.9). Having obtained
(5.9) by exploiting the dual-gauge transformations (5.13), the rest of the
calculations related to the transformation of the polarization
tensor $e^{\mu\nu} (k)$ by the translation subgroup $T(2) \sim W (0, u, v)$
of the Wigner's little group, are exactly the same as (5.10) and (5.11).
It is very interesting to note that (5.11), not only demonstrates the
existence of the gauge transformations (5.4) with the choice of
parameters $\alpha$'s in (5.12), but it also establishes the existence
of the dual-gauge transformations of (5.13) with the following relationships
among the dual-gauge parameters $\Sigma$'s and the parameters of the
translation $T(2)$ subgroup of the Wigner's little group
$$
\begin{array}{lcl}
\Sigma_{1} (k) = - {\displaystyle \frac{i u}{\omega}}\; e^{12} (k),\; \qquad
\Sigma_{2} (k) = - {\displaystyle \frac{i v}{\omega}}\; e^{12} (k),\; \qquad
\Sigma_{3} (k) = - \Sigma_{0} (k),
\end{array} \eqno(5.16)
$$
where the last relation $\Sigma_{3} (k) = - \Sigma_{0} (k)$ emerges from the
dual-gauge transformations (5.13) applied to the transformation
for the polarization tensor component $e^{12} (k)$ (or $e^{21} (k)$). The
expressions for the infinitesimal dual-gauge parameters $\Sigma_{1} (k)$
and $\Sigma_{2} (k)$ in the above equation are derived from the
transformation properties of the polarization tensor components
$e^{01} (k), e^{02} (k), e^{13} (k), e^{23} (k)...$ etc. Moreover, the
``duality'' kind
of transformations of (5.15) are reflected here too if we compare the
relationships obtained in (5.12) and (5.16).

Now let us concentrate briefly
on the other {\it independent} restriction (i.e. $
\varepsilon_{\mu\nu\kappa\zeta}\; e^{\nu\kappa} (k)\; k^\zeta = 0$) that
emerges
from the physicality condition with the dual-BRST charge $Q_{D}$ in (4.21).
From this relation too, we can show that there is only a single degree
of freedom associated with the 4D polarization tensor $e^{\mu\nu} (k)$
for the massless 2-form gauge field $B_{\mu\nu} (x)$.
In contrast to the transversality condition (5.5)
that has been used earlier to derive (5.6),
it can be checked that for $\mu = 0, 1, 2, 3$, we obtain the following
relationships among some of the six-independent components of the 4D
polarization tensor matrix $\{ e^{\mu\nu} (k) \}$
$$
\begin{array}{lcl}
e^{12} = 0,\; \qquad \; e^{01} + e^{13} = 0,\; \qquad \;
e^{02} + e^{23} = 0,
\end{array} \eqno(5.17)
$$
from the dual-transversality relationship
$\varepsilon_{\mu\nu\kappa\zeta}\; e^{\nu\kappa} (k)\; k^\zeta = 0$ where
$k^\mu = (\omega, 0, 0, \omega)^T$.
Thus, the reduced form ($\{ e^{\mu\nu} (k) \}_{(red)})$ of the
polarization tensor is
$$
\begin{array}{lcl}
\{ e^{\mu\nu} (k) \}_{(red)} =
\left ( \begin{array}{cccc}
0 & e^{01} & e^{02} & e^{03} \\
- e^{01} & 0 & 0 & - e^{01} \\
- e^{02} & 0 & 0 & - e^{02}\\
- e^{03} & e^{01} & e^{02} & 0\\
\end{array} \right ).
\end{array} \eqno(5.18)
$$
From the above reduced polarization tensor matrix
$\{ e^{\mu\nu} (k) \}_{(red)}$, one can gauge away
$e^{01} (k)$ and $e^{02} (k)$ either (i)
by the gauge transformations (5.4)
with the choice of and relationship between
the gauge parameters $\alpha$'s as
$$
\begin{array}{lcl}
\alpha^{1} (k) = {\displaystyle \frac{i}{\omega}} \;e^{01} (k),\; \qquad
\alpha^{2}  (k) = {\displaystyle \frac{i}{\omega}} \;e^{02} (k),\; \qquad
\alpha^{0} (k) = \alpha^3 (k),
\end{array} \eqno(5.19)
$$
(where the last restriction emerges from the transformation of the component
$e^{03} (k) = - e^{30} (k)$), or (ii) by choosing the dual-gauge
parameters $\Sigma$'s in transformations (5.13) as
$$
\begin{array}{lcl}
\Sigma_{1} (k) = - {\displaystyle \frac{i}{\omega} }\; e^{02} (k),\; \qquad
\Sigma_{2} (k) = + {\displaystyle \frac{i}{\omega} }\; e^{01} (k),
\end{array} \eqno(5.20)
$$
which is the analogue of (5.14) (with merely a sign difference).
It is clear
that $e^{12} (k) = 0$ due to (5.17). However, as far as the general
dual-gauge transformation (5.13) is concerned, for the validity of the
identity (i.e. $ 0 = 0$)
corresponding to the transformations (5.13) on $e^{12} (k)$,
one has to choose $\Sigma_{3} (k) = - \Sigma_{0} (k)$. Final
reduced version of the polarization tensor (i.e. the analogue of (5.9))
with the restriction $\varepsilon_{\mu\nu\kappa\zeta} \;e^{\nu\kappa} (k)
\; k^\zeta = 0$ is
$$
\begin{array}{lcl}
\{ e^{\mu\nu} (k) \}_{(Red)} = \;e^{03} (k) \;
\left ( \begin{array}{cccc}
0 & 0 & 0 & 1 \\
0 & 0 & 0 & 0 \\
0 & 0 & 0 & 0\\
-1& 0 & 0 & 0\\
\end{array} \right ).
\end{array} \eqno(5.21)
$$
Now we can check the action of the translation subgroup $T(2) \sim
W (0,u,v)$ on the reduced polarization tensor (5.21) in generating the
(dual-)gauge transformations. It can be calculated
explicitly to see that the
analogue of the equation (5.11) is now
$$
\begin{array}{lcl}
 \{ e^{\mu\nu} (k) \} \rightarrow \{ (e^{\mu\nu})^\prime (k) \} =
\{ e^{\mu\nu} (k) \}_{(Red)}
+  e^{03} (k)\;
\left ( \begin{array}{cccc}
0 & - v & - u & 0 \\
u & 0 & 0 & u \\
v & 0 & 0 & v \\
0 & - v & - u & 0\\
\end{array} \right ),
\end{array} \eqno(5.22)
$$
where $\{ e^{\mu\nu} (k) \}_{(Red)}$ is defined in (5.21). This transformation
corresponds to the gauge transformation (5.4) with the following choice of
the infinitesimal gauge parameters $\alpha$'s
$$
\begin{array}{lcl}
\alpha^{1} (k) = {\displaystyle \frac{i u}{\omega}} \;e^{03} (k),\; \qquad
\alpha^{2} (k) = - {\displaystyle \frac{i v}{\omega}} \;e^{03} (k),\; \qquad
\alpha^{3} (k) = \alpha^{0} (k),
\end{array} \eqno(5.23)
$$
where the last relation emerges from the transformation of the polarization
component $e^{03} (k)$ (or $e^{30} (k)$) and the rest of the relationships
 emerge
from the transformations of the components $e^{01} (k), e^{02} (k),
e^{13} (k)..$ etc. It is extremely interesting to point out that
the transformations (5.22) also contain the dual-gauge transformation
of (5.13) for the following choice of the dual-gauge parameters
$$
\begin{array}{lcl}
\Sigma_{1} (k) = - {\displaystyle \frac{i v}{\omega}}\; e^{03} (k),\; \qquad
\Sigma_{2} (k) = - {\displaystyle \frac{i u}{\omega}}\; e^{03} (k).
\end{array} \eqno(5.24)
$$
Here the transformation property of the component $e^{12} (k)$ under the
dual-gauge transformation (5.13) has not been discussed because
$e^{12} (k) = 0$  due to (5.17). However, if one discusses it in the general
framework of dual-gauge transformation (5.13), the corresponding identity
(i.e. $ 0 = 0$) will be satisfied iff $\Sigma_{3} = - \Sigma_{0}$.
It is worth pointing out that (i) here there is no restriction like
the last relationship of (5.16), and (ii) the duality of (5.15) do exist here
too between (5.23) and (5.24).

At this stage, there are a few comments in order.
First, it is unequivocally clear that the conditions
(i) $e^{\mu\nu} (k) k_\nu = - k_\mu e^{\mu\nu} (k) = 0$, and
(ii) $\varepsilon_{\mu\nu\kappa\zeta} e^{\nu\kappa} (k) k^\zeta = 0$
emerging from the physicality criteria (4.21) with the (co-)BRST charges
$(Q_{(D)B}$) imply both the gauge and dual-gauge transformations which
are primarily incorporated in the transformations generated by the
translation $T(2)$
subgroup of the Wigner's little group. Thus, we note that the
(dual-)gauge transformations, (co-)BRST transformations and the transformations
generated by the translation subgroup $T(2)$ of the Wigner's little group are
inter-related. Second,  it can be seen from the
equations (5.5), (5.6) and
(5.9) that the transversality condition and the gauge (or BRST) transformations
imply that only $e^{12} (k)$ is the single degree of freedom left for
the 2-form field $B_{\mu\nu} (x)$. On top of it, if we apply the
dual-transversality condition {\it independently}, we can get rid of
$e^{12} (k)$ as well (cf. (5.17)). Thus, the 2-form field $B_{\mu\nu} (k)$
becomes topological in nature. In fact, it has been shown
that the total (co-)BRST invariant theory, described by (3.12), is
a quasi-topological theory [15] because, in addition to the
topological field $B_{\mu\nu}$, the scalar fields $\phi_1, \phi_2$
and the ghost fields do exist in the theory. Third, we have not
considered here the restrictions emerging from the bosonic conserved charge
$Q_{W}$ because it is automatically implied when both the conditions
due to (co-)BRST charges ($Q_{(D)B}$) are taken into account together in
an independent way.

Let us now come back to the discussion about our comments after
equation (4.15). Exploiting the (anti-)commutators of (4.14)
and (4.15) and invoking the nilpotency
($Q_{(D)B}^2 = 0$) of the (co-)BRST charges ($Q_{(D)B}$), it can be checked
that the following commutation relations
$$
\begin{array}{lcl}
&& [ Q_{B}, \{ Q_{B}, \bar c_\mu^\dagger (k) \} ] = 0,\; \qquad
 [ Q_{B}, \{ Q_{B},  \bar c_\mu (k) \} ] = 0, \nonumber\\
&& [ Q_{D}, \{ Q_{D}, c_\mu^\dagger (k) \} ] = 0,\; \qquad
 [ Q_{D}, \{ Q_{D}, c_\mu (k) \} ] = 0,
\end{array} \eqno(5.25)
$$
are satisfied iff $k^2 = 0, k\cdot \bar c^\dagger = k \cdot \bar c = 0,
k \cdot c^\dagger = k \cdot c = 0$. In fact, the masslessness condition
$ k^2 = 0$ is satisfied because of  our choice of the momentum 4-vectors
$k^\mu = (\omega, 0, 0, \omega)^T, k_{\mu} = (\omega, 0, 0, - \omega)^T$.
However, the latter relations emerge from the condition
$\partial_\mu C^\mu \equiv \partial^\mu C_\mu = (\partial \cdot C) = 0
\Rightarrow k \cdot c^\dagger = k \cdot c = 0$ and condition
$\partial_\mu \bar C^\mu \equiv \partial^\mu \bar C_\mu = (\partial \cdot
\bar C) = 0 \Rightarrow k \cdot \bar c^\dagger = k \cdot \bar c = 0$. We lay
emphasis on the fact that {\it these conditions are met in our whole
discussion.} In fact, it is clear that, in the BRST formalism, the
polarization tensor $e^{\mu\nu} (k)$ transforms as
$$
\begin{array}{lcl}
&&e^{\mu\nu} (k) \rightarrow e^{\mu\nu} (k))^{(db)}
= e^{\mu\nu} (k) + i \varepsilon^{\mu\nu\kappa\zeta} \partial_\kappa
\bar C_\zeta (k),\; \nonumber\\
&&e^{\mu\nu} (k) \rightarrow e^{\mu\nu} (k))^{(b)}
= e^{\mu\nu} (k) + i\; [\; k^\mu C^\nu (k) - k^\nu C^\mu (k) \;],
\end{array} \eqno(5.26)
$$
which are nothing but the generalizations of the (dual-)gauge transformations
to (co-)BRST transformations (5.13) and (5.4)
where $\alpha^\mu (k) \rightarrow C^\mu (k)$ and
$ \Sigma^\mu (k) \rightarrow \bar C^\mu (k)$. It is very clear and transparent
now that the last relationships of the (dual-)gauge transformations
(5.16) and (5.12) are replaced by the vector (anti-)ghost fields of the
(co-)BRST transformations as
$$
\begin{array}{lcl}
\bar C_{3} (k) = - \bar C_{0} (k), \;\;\qquad \;\; C^3 (k) = C^0 (k),
\end{array} \eqno(5.27)
$$
which very convincingly satisfy $ k \cdot C (k) = k \cdot \bar C (k) = 0$
implying $k \cdot c^\dagger = k \cdot c = 0, \; k \cdot \bar c^\dagger =
k \cdot \bar c = 0$ with our choice of the momentum vectors
as $ k^\mu = (\omega, 0, 0, \omega)^T, k_\mu = (\omega, 0, 0, - \omega)^T$.
The above conclusions can be drawn from the (dual-)gauge transformations
(5.24) and (5.23) as well. For the proof of $ \bar C_{3} (k)
= - \bar C_{0} (k)$, it will be worthwhile to go through the discussions
after equation (5.24). Thus, it is obvious that all the
crucial comments after
(4.15) are justified and there is nothing in the theory that has been
imposed from outside.

Now we would like to end this section with a few brief comments about the
connection between the (dual-)gauge transformed states (that are also
connected with the transformation on the polarization tensor $e^{\mu\nu} (k)$
by the translation $T(2)$ subgroup of the Wigner's little group) in the
QHSS and the BRST cohomology w.r.t. the
conserved and nilpotent (co-)BRST charges. It is evident that a single
particle quantum state of the 2-form field (SPQS) with the polarization
tensor $e^{\mu\nu} (k)$ can  be created from the vacuum by the application
of the creation operator $b_{\mu\nu}^\dagger (k)$
(i.e. $e^{\mu\nu} (k) b_{\mu\nu}^\dagger (k)\;|vacm>$) as given by (4.16).
Exploiting the (anti-)commutation relations of (4.14), it can be checked that
$$
\begin{array}{lcl}
&&- 2\; [\; Q_{B}, (c^\mu)^\dagger (k) \bar c_\mu^\dagger (k) \; ]
= - 2 \;\{ Q_{B}, (c^\mu)^\dagger (k) \} \;\bar c_\mu^\dagger (k)
+  2 (c^\mu)^\dagger (k)\; \{ Q_{B},  \bar c_\mu^\dagger (k) \} \nonumber\\
&&\equiv 2 b^\dagger (k)\; [k \cdot \bar c^\dagger (k)] + 2 \;
[ k \cdot c^\dagger (k) ]\; f_{1}^\dagger (k) - 2 \;
[k^\nu (c^\mu)^\dagger (k) ] \;b_{\mu\nu}^\dagger (k).
\end{array} \eqno(5.28)
$$
Applying our observation that $k \cdot \bar c^\dagger = k \cdot c^\dagger = 0$,
it is straightforward to see that the above commutator, ultimately, reduces
to the following expression (with $b_{\mu\nu}^\dagger (k) =
- b_{\nu\mu}^\dagger (k)$)
$$
\begin{array}{lcl}
- 2\; [\; Q_{B}, (c^\mu)^\dagger (k) \bar c_\mu^\dagger (k) \; ]
= + [ \; k^\mu (c^\nu)^\dagger (k) - k^\mu (c^\mu)^\dagger (k) \; ]\;
b_{\mu\nu}^\dagger (k).
\end{array} \eqno(5.29)
$$
The physical implication of the above equation emerges when it is applied
on the physical vacuum state (i.e. $ Q_{B} \; |vacm> = 0$) of the theory
as given below
$$
\begin{array}{lcl}
&& [ \; k^\mu (c^\nu)^\dagger (k) - k^\mu (c^\mu)^\dagger (k) \; ]\;
b_{\mu\nu}^\dagger (k)\; | vacm > =
- 2\; [\; Q_{B}, (c^\mu)^\dagger (k) \bar c_\mu^\dagger (k) \; ] \; | vacm>
\nonumber\\
&&\equiv Q_{B} \bigl ( -2
[\; (c^\mu)^\dagger (k) \bar c_\mu^\dagger (k) \; ] \bigr )\; | vacm>.
\end{array} \eqno(5.30)
$$
It is lucid and clear that the above state is a BRST exact state and, hence,
in the language of HDT and BRST cohomology, it is a trivial state. Taking the
sum of the equations (4.16) and (5.30) (with a factor of $i$), we obtain
the following relationship between the gauge (i.e. BRST) transformed SPQS
and the original SPQS (4.16)
$$
\begin{array}{lcl}
&&\Bigl (\; e^{\mu\nu} (k)
+ i\; [ \; k^\mu (c^\nu)^\dagger (k) - k^\mu (c^\mu)^\dagger (k) \; ]\; \Bigr )
\;b_{\mu\nu}^\dagger (k)\; | vacm > \nonumber\\
&&= e^{\mu\nu} (k)\; b_{\mu\nu}^\dagger (k)\; |vacm>
+ Q_{B} \bigl ( \;- 2 i [ (c^\mu)^\dagger (k) \bar c_\mu^\dagger (k) \; ]
\;\bigr ) \; | vacm>.
\end{array} \eqno(5.31)
$$
This establishes the fact that a gauge (or BRST) transformed state in
the quantum Hilbert space (that is connected with the transformation on
the polarization tensor $e^{\mu\nu} (k)$ by the translation subgroup
$T(2)$ of the Wigner's little group) is the sum of the original SPQS plus
a BRST exact state. In more sophisticated language, the gauge (or BRST)
transformed SPQS and the original SPQS belong to the same cohomology
class w.r.t. the conserved ($\dot Q_{B} = 0$)
and nilpotent ($Q_{B}^2 = 0$) BRST charge $Q_{B}$.
Similarly, it can be checked, exploiting the (anti-)commutators of (4.15),
that the dual-gauge (or co-BRST) transformed SPQS
and the original SPQS in the quantum Hilbert space of states are related to
each-other as
$$
\begin{array}{lcl}
&&\Bigl (\; e^{\mu\nu} (k)
+ i\; [ \; \varepsilon^{\mu\nu\sigma\zeta}\; k_\sigma \bar c_\zeta^\dagger (k)
\; ]\; \Bigr )
\;b_{\mu\nu}^\dagger (k)\; | vacm > = e^{\mu\nu} (k)\;
b_{\mu\nu}^\dagger (k)\; |vacm>
\nonumber\\
&&+ Q_{D} \bigl ( \;- 2 i [ (\bar c^\mu)^\dagger (k) c_\mu^\dagger (k) \; ]
\;\bigr ) \; | vacm>,
\end{array} \eqno(5.32)
$$
where we have used the following commutation relationship
$$
\begin{array}{lcl}
&&- 2\; [\; Q_{D}, (\bar c^\mu)^\dagger (k)  c_\mu^\dagger (k) \; ]
= - 2 \;\{ Q_{D}, (\bar c^\mu)^\dagger (k) \} \;c_\mu^\dagger (k)
+  2 (\bar c^\mu)^\dagger (k)\; \{ Q_{D},  c_\mu^\dagger (k) \} \nonumber\\
&&\equiv - 2 \bar b^\dagger (k)\; [k \cdot  c^\dagger (k)] + 2 \;
[ k \cdot \bar c^\dagger (k) ]\; f_{2}^\dagger (k) +
\varepsilon_{\mu\nu\kappa\sigma}\;k^\mu\;
(\bar c^\nu)^\dagger \;(b^{\kappa\sigma})^\dagger (k).
\end{array} \eqno(5.33)
$$
This demonstrates that a dual-gauge (or co-BRST) transformed SPQS and
the original SPQS belong to the same cohomology class w.r.t. the conserved
($\dot Q_{D} = 0$) and nilpotent ($Q_{D}^2 = 0$) co-BRST charge $Q_{D}$.
Thus, it is obvious that if we take into account the following three
basic ideas: (i) the BRST cohomology w.r.t. (co-)BRST charges (ii)
the HDT applied to the states of the quantum Hilbert space, and (iii) the
choice of the physical states (as well as the vacuum) to be the harmonic
states of the HDT, it can be proven explicitly that
the (dual-)gauge (or (co-)BRST) transformed SPQS
(that corresponds to the transformations on $e^{\mu\nu} (k)$ by the
translation subgroup $T(2)$ of the Wigner's little group) and the original
SPQS belong to the same cohomology class w.r.t. (co-)BRST charges $Q_{(D)B}$.

As commented after the equation (5.3), it is unequivocally clear that the
Wigner's little group for any arbitrary 2D gauge theory is trivial. In fact,
the matrix representation for it becomes identity matrix (cf. (5.1)). Thus,
the (dual-)gauge transformations, discussed in section 2, for the free 2D
Abelian gauge theory can not be described in the framework of Wigner's little
group. However, it can be discussed in the framework of BRST cohomology
based on the constraint analysis of the theory. In fact, physically
there is nothing in the free 2D Abelian gauge theory because
of the fact that both the degrees of freedom of the photon can be gauged away
by the (dual-)gauge (or (co-)BRST) transformations. In
other words, both the components of the polarization vector $e_\mu (k)$
in 2D can be gauged away by the (dual-)gauge
(or (co-)BRST) transformations. Thus, the
arguments of the translation subgroup $T(2)$
of the Wigner's little group does not work for 2D free Abelian gauge theory.\\

\noindent
{\bf 6 Conclusions}\\

\noindent
In our present investigation, we have demonstrated
an interesting connection between
(i) the (dual-)gauge transformations on the polarization tensor $e^{\mu\nu} (k)
= - e^{\nu\mu} (k)$ of the 4D free Abelian 2-form gauge field
generated by the first-class constraints of the theory, and (ii) similar
transformations generated by the (Abelian invariant) two-parameter
translation subgroup $T(2)$ of the Wigner's little group. Both the above
transformations are shown to be equivalent for certain specific
relationships among the (dual-)gauge parameters
of the internal symmetry transformation group  and the parameters
of the translation subgroup $T(2)$ of the Wigner's little group
(which does not transform the momentum vector
$k_\mu$ of the massless gauge particle).
The latter parameters characterize the Euclidean plane $(xy)$ which is
perpendicular to the $z$-direction of the propagation of the massless
($ k^2 = 0$) gauge particle. In the framework of the BRST cohomology and HDT
applied to the QHSS, it is established in our present endeavour
that the (dual-)gauge (or (co-)BRST)
transformed states in the QHSS are the sum of the
original (untransformed) states plus the (co-)BRST exact states. Thus, it
becomes crystal clear that {\it the changes} in the original state
due to the (dual-)gauge
(or (co-)BRST) transformations correspond to the cohomologically
{\it trivial} states as they satisfy trivially the
physicality condition $ Q_{B} | phys> = 0, Q_{D} |phys> = 0,
Q_{W} |phys> = 0$ for our choice of the physical
states to be the harmonic states. This happens primarily due to
the nilpotency ($Q_{(D)B}^2 = 0$) of the (co-)BRST charges
$(Q_{(D)B}$) and the definition of the bosonic charge
$Q_{W} = Q_{B} Q_{D} + Q_{D} Q_{B}$ (which turns out to be the analogue of the
Laplacian operator). These statements can be proven in terms of the nilpotent
and conserved anti-BRST charge
$(Q_{AB})$ and anti-co-BRST charge $(Q_{AD})$ as well.

It is interesting to point out that, in the framework of BRST formalism,
the equation of motion $\Box B_{\mu\nu} = 0$ for the 2-form gauge field
is such that the masslessness condition ($k^2 = 0$) is implied automatically
unlike the discussions in [36] where the massive and massless cases for the
above field are considered separately. In fact, in [36], the Lagrangian density
is not gauge-fixed to start with and it contains the kinetic energy term
(${\cal L}_{0} = \frac{1}{12} H^{\mu\nu\kappa} H_{\mu\nu\kappa}$) only. This
is why the transversality condition $k_\mu\; e^{\mu\nu} (k) = -
e^{\mu\nu} (k)\; k_\nu = 0$ emerges from the equation of motion when the
masslessness condition ($k^2 = 0$) is imposed from outside by hand. This is
not the case in our discussion because it is based on the BRST formalism where
the Lagrangian density is gauge-fixed right from the beginning. In fact, for
our description in the framework of the BRST formalism, the transversality
condition (i.e. $k_\mu e^{\mu\nu} (k)
= - e^{\mu\nu} (k) k_\nu = 0$) and the dual-transversality condition
(i.e. $\varepsilon_{\mu\nu\sigma\zeta} e^{\nu\sigma} (k) k^\zeta = 0$)
emerge from the physicality criteria w.r.t. the conserved and
nilpotent (co-)BRST charges (cf. (4.21)). It is extremely gratifying
to point out that the transformed polarization matrices
(5.11) and (5.22) due to $T(2)$ subgroup of the Wigner's little group
contain both the gauge and the dual-gauge transformations (5.4) and
(5.13) for the specific choices and inter-relationships
among some of the (dual-)gauge transformation parameters. In fact, the
latter transformation parameters are chosen in terms of the parameters of the
translation $T(2)$ subgroup of the Wigner's little group in a
certain specific way (cf. (5.12), (5.16), (5.23), (5.24))
for the validity of our above claim. These statements can be recast
and re-expressed in the language of  BRST formalism where the
(dual-)gauge transformation parameters are replaced by the anti-commuting
(anti-)ghost fields.
Another interesting feature, connected with the equations of motion for the
fermionic vector (anti-)ghost fields
(i.e. $\Box C_\mu = \frac{3}{2} \partial_\mu (\partial \cdot C),
\Box \bar C_\mu = \frac{3}{2} \partial_\mu (\partial \cdot \bar C)$) {\it is
the normal mode expansions in (4.13).}
It will be noted that the
expansion in (4.13) can be true iff $\Box C_\mu = \Box \bar C_\mu = 0$
which imply that $(\partial \cdot C) = (\partial \cdot \bar C) = 0$. These
conditions in the momentum phase space will correspond to $k^2 = 0,\;
k \cdot C = k \cdot \bar C = 0$. It is nice to point out that these
conditions are met throughout our discussions because of the choice of the
4-momentum vectors $k^\mu = (\omega, 0, 0, \omega)^T, k_{\mu} =
(\omega, 0, 0, - \omega)^T$ and the conditions $ \bar C_3 = - \bar C_0,
C^3 = C^0$ (cf. (5.27)) on the (anti-)ghost fields
which emerge due to the {\it equivalence}
between the (dual-)gauge (or (co-)BRST) transformations
on the polarization tensor $e^{\mu\nu} (k)$ and such
type of transformations generated
by the translational subgroup $T(2)$ of the Wigner's little group. Rest of the
normal mode expansions in (4.13) are consistent with the equations of motion
emerging from the Lagrangian density (3.12) that respects the on-shell
nilpotent (co-)BRST symmetries. Thus, {\it the Wigner's little group plays a
very decisive
and crucial role in the correctness of the normal mode expansion in (4.13).}

For the 2D free Abelian one-form gauge theory, it is clear that the Wigner's
little group becomes trivial and it generates {\it no} transformation on the
polarization vector $e_\mu (k)$. By contrast, in the framework of the BRST
cohomology, the (dual-)gauge transformations on $e_\mu (k)$ can be
discussed elegantly which finally imply that this theory belongs to a
new class of topological field theory (see, e.g., [10] for details). As far
as the degrees of freedom count on $e^{\mu\nu} (k)$ for the 2-form gauge
theory is concerned, it turns out that all the components of $e^{\mu\nu} (k)$
can be gauged away by exploiting the (co-)BRST symmetries {\it together},
treating them in an {\it independent} way. However, it has been shown [15]
that the total (co-)BRST invariant Lagrangian density (3.12) does not represent an exact topological field theory. Rather, it presents an example of a
quasi-topological field theory in the flat 4D spacetime (see, e.g., [15] for
details) because the additional scalar fields $\phi_1, \phi_2$ and
the ghost fields do exist in the theory besides the topological field
$B_{\mu\nu}$. It would be nice to generalize
our present work to the discussion of the higher rank anti-symmetric tensor
fields in 4D and more than 4D of spacetime following the method adopted in
[43]. Another direction that could be pursued is the discussion of the
(dual-)gauge (or (co-)BRST) type symmetries for the case of the linearized
gravity theory which has been recently studied in the framework of the
Wigner's little group [44]. These are some of the issues that are under
investigation and our results would be reported elsewhere in our
future publications [45].\\

\noindent
{\bf Note Added in the Proof}:
After our paper was accepted for publication, we came to
know about a paper: H. Hata, T. Kugo and N. Ohta {\it Nucl. Phys.}
{\bf 178}, 527 (1981) where BRST analysis has been performed,
albeit in a different context, for the 2-form gauge field.

\end{document}